\documentclass[english,aps,nofootinbib,prd,preprint,showpacs,fleqn]{revtex4}
\usepackage[T1]{fontenc}
\usepackage[latin1]{inputenc}
\usepackage{subfigure}
\usepackage{graphicx}
\usepackage{amssymb}

\makeatletter

\providecommand{\LyX}{L\kern-.1667em\lower.25em\hbox{Y}\kern-.125emX\@}

\def\D0{D\0~}

\def\STAR{{\sc Star }}
\def\PHENIX{{\sc Phenix }}

\usepackage{babel}
\makeatother
\begin{document}

\newcommand{\wh}[1]{\widehat{#1}}

\newcommand{\wt}[1]{\widetilde{#1}}

\newcommand{\MSbar}{\overline{MS}}

\newcommand{\A}{{\mathcal{A}}}

\newcommand{\eps}{\varepsilon }

\newcommand{\B}{{\mathcal{B}}}

\newcommand{\C}{{\mathcal{C}}}

\renewcommand{\floatpagefraction}{1} \renewcommand{\S}{{\cal S}}\renewcommand\vec[1]{{\bf #1}}

\preprint{hep-ph/0304002}


\title{Single-spin physics with weak bosons at RHIC}

\author{P.M. Nadolsky\protect$^{1}$}

\email{nadolsky@mail.physics.smu.edu}

\author{C.-P. Yuan\protect$^{2}$}

\email{yuan@pa.msu.edu}

\affiliation{$^{1}$Department of Physics, Southern Methodist University, \\
Dallas, Texas 75275-0175, U.S.A.\\
$^{2}$Department of Physics \& Astronomy, Michigan State University,
East Lansing, Michigan 48824, U.S.A.}

\date{31st March, 2003}

\begin{abstract}
In order to measure spin-dependent parton distributions in production
of $W^{\pm }$ bosons using the non-hermetic detectors of the Relativistic
Heavy Ion Collider, an accurate model for distributions of charged
leptons from the $W$ boson decay is needed. We discuss the predictions
for production and decay of the $W$ bosons based on a calculation
for resummation of large logarithmic contributions originating from
multiple soft gluon radiation. We compare the predictions for the
spin-dependent resummed cross sections with the corresponding next-to-leading
order results. We show that the lepton-level asymmetries can be reliably
predicted by the resummation calculation and directly compared to
the experimental data. A program for the numerical analysis of such
cross sections in $\gamma ^{*},\, W^{\pm },$ and $Z^{0}$ boson production
is also presented. 
\end{abstract}
\maketitle

\section{\label{sec:Intro}Introduction}

The measurement of longitudinal spin asymmetries in production of
$W^{\pm }$ bosons at the Relativistic Heavy Ion Collider (RHIC) will
provide an essential probe for spin-dependent quark distributions
at high momentum scales $Q^{2}$ \cite{RHICOverview}. At a $pp$
center-of-mass energy $\sqrt{s}=500$ GeV, about $1.3\times 10^{6}$
$W^{+}$and $W^{-}$ bosons will be produced by the time the integrated
luminosity reaches $800\mbox {\, pb}^{-1}$. Due to the parity violation
in the $q\bar{q}W$ coupling, this process permits non-vanishing single-spin
asymmetries $A_{L}(\xi )$, defined for any kinematical variable $\xi $
as\begin{equation}
A_{L}(\xi )\equiv \frac{\frac{d\sigma (p^{\rightarrow }p\rightarrow WX)}{d\xi }-\frac{d\sigma (p^{\leftarrow }p\rightarrow WX)}{d\xi }}{\frac{d\sigma (p^{\rightarrow }p\rightarrow WX)}{d\xi }+\frac{d\sigma (p^{\leftarrow }p\rightarrow WX)}{d\xi }}.\label{ALxi}\end{equation}
 The Born-level expression for the asymmetry $A_{L}(y_{W})$ with
respect to the rapidity $y_{W}$ of the $W$ boson is particularly
simple if the absolute value of $y_{W}$ is large. In that case, the
Born-level $A_{L}(y_{W})$ reduces to the ratio $\Delta q(x)/q(x)$
of the polarized and unpolarized parton distribution functions \cite{Bourrely:1994sc,Bourrely:1995fw,Nadolsky:1995nf}.
Furthermore, $A_{L}(y_{W})$ tests the flavor dependence of quark
polarizations. 

The original method for extracting the spin-dependent quark distributions
out of the RHIC $W$ boson data is based on the direct reconstruction
of the asymmetry $A_{L}(y_{W})$ \cite{RHICOverview,Bland:1999gb}.
Unfortunately, such reconstruction is obstructed by specifics of the
detection of $W^{\pm }$ bosons at RHIC. First, neither \PHENIX \cite{PHENIX}
nor \STAR \cite{STAR} detector at RHIC is hermetic. Therefore, it
is not possible at those detectors to monitor the energy balance in
particle reactions, even in the transverse direction with respect
to the beams. Due to the lack of information about the missing energy
carried by the neutrino, the determination of $y_{W}$ is in general
ambiguous and depends on the assumptions about the dynamics of the
process. Second, due to the correlation between the spins of the initial-state
quarks and final-state leptons, the measured value of $A_{L}(y_{W})$
is strongly sensitive to the experimental cuts imposed on the observed
charged lepton. Hence, unless a theory calculation exists to reliably
predict the distributions of the leptons from the $W$ boson decay,
the spin-dependent quark distribution functions cannot be determined
with acceptable accuracy. 

In weak boson production at the Fermilab Tevatron $p\bar{p}$ collider,
the mass $M_{W}$ of the $W$ boson has been measured to a great precision
\cite{Affolder:2000bp,Abazov:2002bu}. The information on $M_{W}$
is extracted out of the lepton-level distributions $d\sigma /m_{T}$,
$d\sigma /dp_{Te}$, or $d\sigma /dp_{T\nu }$, where $p_{Te}\equiv |\vec{p}_{Te}|$
and $p_{T\nu }\equiv |\vec{p}_{T\nu }|$ are the transverse momenta
of the electron and neutrino, and $m_{T}\equiv \sqrt{2p_{Te}p_{T\nu }-2\vec{p}_{Te}\cdot \vec{p}_{T\nu }}$
is the transverse mass of the lepton pair. The value of $M_{W}$ can
be determined from the analysis of the position and shape of the Jacobian
peaks in each of these three distributions. For most of the events
contributing near the Jacobian peak, the transverse momentum $q_{T}$
of the $W$ boson is very small, of order of a few GeV. Therefore,
at each order of the perturbative QCD calculation, there are large
logarithmic contributions of the form $\alpha _{S}^{n}q_{T}^{-2}\ln ^{m}\/ (M_{W}^{2}/q_{T}^{2})$,
where $n=1,2,\dots $ and $m=0,1,\dots ,2n-1$. These large logarithmic
contributions can be summed to all orders in $\alpha _{S}$ by applying
the transverse momentum resummation formalism formulated in the impact
parameter space by Collins, Soper, and Sterman (CSS) \cite{CSS}.
Without such a resummation calculation performed at the decay lepton
level \cite{CPCsaba}, it would not be possible to determine $M_{W}$
from the experimental data. This is because at hadron colliders the
longitudinal momentum of the neutrino from the $W$ boson decay is
not observed by the detectors; hence, the invariant mass of the $W$
boson cannot be measured directly. Instead, the determination of the
$W$ boson mass relies on the analysis of the shape of the Jacobian
peaks in the lepton-level distributions sensitive to $q_{T}$. Again,
without a resummation calculation at the secondary lepton level, it
would not be possible to predict the shape of these distributions
accurately enough to determine $M_{W}$. We refer the reader to Ref.~\cite{CPCsaba}
for a detailed discussion of the phenomenology of the $W$ boson physics
at the Tevatron collider. 

Similarly, at RHIC, a resummation calculation is needed to reliably
predict the distribution of the secondary leptons from the $W$ boson
decay. As we already mentioned, the reason is the necessity to deduce
spin-dependent parton luminosities based on the observation of the
secondary charged lepton only. For RHIC purposes, the resummation
calculation at the lepton level has to be modified to include the
spin dependence, which has been performed in Ref.~\cite{PolWTheory}.
In this paper, we will apply the calculation in Ref.~\cite{PolWTheory}
to model single-spin asymmetries of $W$ boson production in realistic
RHIC conditions. We put the emphasis on the discussion of the asymmetries
$A_{L}(y_{\ell })$ and $A_{L}(p_{T\ell })$ for the distributions
in the rapidity $y_{\ell }$ and transverse momentum $p_{T\ell }$
of the observed charged lepton. We will argue that these directly
observed asymmetries provide a viable alternative to the commonly
discussed asymmetry $A_{L}(y_{W})$. 

The paper is organized as follows. In Section \ref{sec:FormalismOutline},
we briefly summarize the resummation formalism used for this study.
A next-to-leading order (NLO) calculation is also presented for comparison
to the resummation calculation. Since good understanding of spin-averaged
cross sections is needed to derive polarized PDFs from the spin asymmetries,
Section~\ref{sec:Unpolarized} discusses the uncertainties due to
the imperfect knowledge of unpolarized parton distributions, as well
as the potential for RHIC experiments to reduce these uncertainties.
Section~\ref{sec:Reconstruction-of-spin} demonstrates that a correct
model of multiple parton radiation is important to describe the differential
distributions of the final-state leptons. We then employ the resummation
calculation to realistically evaluate the feasibility of the reconstruction
of the asymmetry $A_{L}(y_{W})$ from the observed charged lepton
data. Section~\ref{sec:Lepton-level-asymmetries} shows that the
spin asymmetries for the distributions of the observed secondary leptons
provide an attractive alternative to the asymmetry $A_{L}(y_{W})$:
they can be measured directly and discriminate efficiently between
the different PDF sets. We also show that the RHIC experiments can
explore the spin dependence of the nonperturbative contributions from
the multiple parton radiation. The main findings of this paper are recapped
in the conclusion.

\section{\label{sec:FormalismOutline} Theoretical essentials}

The primary goal of this paper is to discuss production of $W^{\pm }$
bosons, which will be used at RHIC to obtain information
about the longitudinally polarized parton distribution functions (PDFs)
in the quark sector. Due to the unique feature of the maximal parity
violation in the $Wq\bar{q}$ coupling, the parton-level cross sections
for spin-dependent $W$ boson production have pronounced single-spin
asymmetries. This feature gives $W$ boson production an edge on parity-conserving
processes, in which non-trivial hadronic dynamics can be probed only
through the measurement of more complex double-spin asymmetries. 

Denote the unpolarized distribution of a parton $a$ in a nucleon
$A$ as $f_{a/A}(\xi ,\mu _{F})$ and the longitudinally polarized
distribution of $a$ in the nucleon $A$ as $\Delta f_{a/A}(\xi ,\mu _{F})$.
Then, \begin{eqnarray}
f_{a/A}(\xi ,\mu _{F}) & \equiv  & f_{+/+}(\xi ,\mu _{F})+f_{-/+}(\xi ,\mu _{F}),\\
\Delta f_{a/A}(\xi ,\mu _{F}) & \equiv  & f_{+/+}(\xi ,\mu _{F})-f_{-/+}(\xi ,\mu _{F}),\label{fP}
\end{eqnarray}
 where $f_{h_{a}/h_{A}}(\xi _{a},\mu _{F})$ is a helicity-dependent
parton distribution function, \textit{\emph{i.e.}}, a probability
of finding a parton $a$ with the momentum $p_{a}^{\mu }=\xi _{a}p_{A}^{\mu }$
and helicity $h_{a}$ in a hadron $A$ with the momentum $p_{A}^{\mu }$
and helicity $h_{A}.$ The PDFs depend on the factorization scale
$\mu _{F}$, which in our calculation is assumed to coincide with
the QCD renormalization scale. To find the polarized PDFs from an
experiment with one longitudinally polarized beam, one measures the
unpolarized cross section, \begin{equation}
\frac{d\sigma }{d\xi }=\frac{1}{2}\left(\frac{d\sigma (p^{\rightarrow }p\rightarrow WX)}{d\xi }+\frac{d\sigma (p^{\leftarrow }p\rightarrow WX)}{d\xi }\right),\end{equation}
as well as the single-spin cross section,\begin{equation}
\frac{d\Delta _{L}\sigma }{d\xi }=\frac{1}{2}\left(\frac{d\sigma (p^{\rightarrow }p\rightarrow WX)}{d\xi }-\frac{d\sigma (p^{\leftarrow }p\rightarrow WX)}{d\xi }\right).\label{sigmaL}\end{equation}
Here $\xi $ denotes any kinematic variable, such as the rapidity
$y_{W}$ of the $W$ boson. The single-spin asymmetry is defined as
\begin{equation}
A_{L}(\xi )\equiv \frac{d\Delta _{L}\sigma /d\xi }{d\sigma /d\xi }.\end{equation}

In Ref.~\cite{PolWTheory}, we presented fully differential unpolarized
cross sections $d\sigma /d\xi $, single-spin cross sections $d\Delta _{L}\sigma /d\xi ,$
and double-spin cross sections $d\Delta _{LL}\sigma /d\xi $ for production
and decay of $\gamma ^{*},$$W^{\pm },$ and $Z^{0}$ bosons. To reliably
predict the rate at any point of the momentum phase space, these cross
sections include the all-order sum of large soft and collinear logarithms,
which dominate at small transverse momentum $q_{T}$ of the vector
boson. This all-order sum is combined, without double counting, with
the finite-order ${\mathcal{O}}(\alpha _{S})$ cross section, which
dominates at large $q_{T}$. The single-spin resummed cross sections
in the narrow width approximation were first obtained in Ref.~\cite{Weber2}.
The main result of Ref.~\cite{PolWTheory} is the (more involved)
derivation of the resummed cross sections that also accounts for the
decay of the vector bosons and angular distributions of the final-state
leptons. Our result was derived in the $\overline{MS}$ factorization
scheme \cite{Mertig:1996ny,Vogelsang:1996vh,VogelsangTwoLoop}; therefore,
it is fully compatible with the $\overline{MS}$ parton distributions. 

The resummed cross sections from Ref.~\cite{PolWTheory} are implemented
in the numerical resummation program \textsc{Legacy} and Monte-Carlo
integration program \textsc{RhicBos,}%
\footnote{The Fortran code and input grids for \textsc{RhicBos} can be downloaded
from http://hep.pa.msu.edu/\textasciitilde{}nadolsky/RhicBos/. %
} \textsc{}which were used to produce the results discussed in this
paper. \textsc{Legacy} is a C++/Fortran program that quickly and accurately
generates the resummed cross sections on a grid of points in the kinematical
phase space. \textsc{RhicBos} reads in the cross section grids, performs
their multi-dimensional integration, and produces the output that
is optimized for RHIC specifics. Since the finite-order cross section
is characterized by a smaller theoretical uncertainty at $q_{T}\gtrsim Q$,
\textsc{RhicBos} switches from the resummed cross section ($W+Y$)
to the finite-order cross section at the point where the resummation
cross section becomes smaller than the finite-order cross section
at $q_{T}$ above the position of the maximum in $d\sigma /dq_{T}$.
To compare the resummation calculation to the finite-order approach,
we have also calculated the finite-order ${\mathcal{O}}(\alpha _{S})$
cross section (next-to-leading order, or NLO, cross section) with
the help of a phase space slicing method. This method introduces a
separation transverse momentum scale $q_{T}^{sep}$, below which the
integral of $d\sigma /dq_{T}^{2}$ over the region $0\leq q_{T}^{2}\leq (q_{T}^{sep})^{2}$
is calculated analytically, using the small-$q_{T}$ approximation
in Eq.~(78) of Ref.~\cite{PolWTheory}. The integral over the region
$q_{T}^{2}\geq (q_{T}^{sep})^{2}$ is integrated numerically using
the exact ${\mathcal{O}}(\alpha _{S})$ cross section. As we will
show later, in the NLO approach, predictions for several important
distributions depend strongly on the value of the auxiliary parameter
$q_{T}^{sep}$, so that the resummed cross sections must be used in
order to model those distributions. Some other distributions, such
as $d\sigma /dQ^{2}$, are not sensitive to $q_{T}^{sep}$ (up to
higher-order corrections), so that the resummation and NLO calculations
give close results for such distributions. 

The electroweak parameters in this paper were obtained in the on-shell
scheme according to the procedure described in Section II.B of Ref.~\cite{PolWTheory}
for the Fermi constant $G_{F}=1.16639\times 10^{-5}\mbox {\, GeV}^{-2}$,
$W$ boson mass $M_{W}=80.419$ GeV, and $Z^{0}$ boson mass $M_{Z}=91.187$
GeV. The width $\Gamma _{W}$ of the $W$ boson was evaluated as $\Gamma _{W}=3G_{F}M_{W}^{3}/(2\pi \sqrt{2})$.
The finite-order and asymptotic cross sections were calculated for
the QCD factorization scale $\mu _{F}=Q$. The resummed cross section
(see Eqs.~(79) and (80) in Ref.~\cite{PolWTheory}) was calculated
for the canonical values of the scale parameters, $C_{1}=C_{3}=2e^{-\gamma _{E}}$,
where $\gamma _{E}\approx 0.577$ is the Euler constant, and $C_{2}=1$.
Unless stated otherwise, the nonperturbative Sudakov factor in the
resummed cross sections was evaluated by using the recent Gaussian
parametrization \cite{Landry:2002ix} derived from the global analysis
of transverse momentum distributions in unpolarized vector boson production.
The projected statistical errors $\delta A_{L}$ in the measurement
of the single-spin asymmetries were estimated according to Eq.~(13)
in Ref.~\cite{RHICOverview}: \begin{equation}
\delta A_{L}=\sqrt{\frac{1}{NP^{2}}-\frac{1}{N}A_{L}^{2}},\label{deltaAL}\end{equation}
where $N={\mathcal{L}}\sigma $ is the number of the events for the
given luminosity ${\mathcal{L}}$ and cross section $\sigma $. Unless
stated otherwise, the assumed integrated luminosity is ${\mathcal{L}}=800\mbox {\, pb}^{-1}$,
and polarization $P$ of the beam $A$ is $70\%$.

\section{\label{sec:Unpolarized} Physics potential in unpolarized $pp$ collisions}

\begin{table}[p]

\caption{\label{table:xsect} The unpolarized cross sections (in pb) and numbers
of events $N$ for the processes $pp\rightarrow (W^{\pm }\rightarrow \ell \nu _{\ell })X$
and $pp\rightarrow (Z^{0}\rightarrow \ell \bar{\ell })X$ at RHIC.
Here $\ell $ and $\nu _{\ell }$ are leptons of one lepton generation,
and the decay branching ratios were calculated to be $\mbox {Br}\left(W^{\pm }\rightarrow \ell \nu _{\ell }\right)=0.11,$
$\mbox {Br}\left(Z^{0}\rightarrow \ell \bar{\ell }\right)=0.034$.
The numbers in parentheses show the relative uncertainties. Also shown
are the Born-level momentum fractions $x=Q/\sqrt{s}$ corresponding
to $y_{W}=0$ (or $y_{Z}=0$). The uncertainties $\delta \sigma _{PDF}$
in the cross sections $\sigma $ are due to the uncertainties in the
parton distribution functions estimated according to the CTEQ6 analysis
\cite{Pumplin:2002vw,Stump:2003yu}. The uncertainties in $N$ are
purely statistical.}

\begin{tabular}{|c|c|c|c|}
\cline{3-3} \cline{4-4} 
\multicolumn{2}{c|}{}&
~~~~~$\sqrt{s}=200$ GeV~~~~~&
~~~~~$\sqrt{s}=500$GeV~~~~~\\
\multicolumn{2}{c|}{}&
${\mathcal{L}}=320\mbox {\, pb}^{-1}$&
${\mathcal{L}}=800\mbox {\, pb}^{-1}$\\
\hline 
&
$\left.x\right|_{y_{W}=0}$&
$0.4$&
$0.16$\\
~$W^{+}$~&
~~$\sigma \pm \delta \sigma _{PDF}$$\left(\frac{\delta \sigma _{PDF}}{\sigma }\right)$~~&
$1.38\pm 0.34\, (0.25)$&
$124\pm 9\, (0.07)$\\
&
$N\pm \sqrt{N}$$(1/\sqrt{N})$&
$440\pm 20\, (0.05)$&
$99200\pm 300\, (0.003)$\\
\hline
&
$\left.x\right|_{y_{W}=0}$&
$0.4$&
$0.16$\\
$W^{-}$&
$\sigma \pm \delta \sigma _{PDF}$$\left(\frac{\delta \sigma _{PDF}}{\sigma }\right)$&
$0.43\pm 0.12\, (0.27)$&
$41\pm 4\, (0.10)$\\
&
$N\pm \sqrt{N}$$(1/\sqrt{N})$&
$142\pm 12\, (0.09)$&
$32800\pm 200\, (0.006)$\\
\hline 
&
$\left.x\right|_{y_{Z}=0}$&
$0.46$&
$0.18$\\
$Z^{0}$&
$\sigma \pm \delta \sigma _{PDF}$$\left(\frac{\delta \sigma _{PDF}}{\sigma }\right)$&
$0.07\pm 0.02\, (0.26)$&
$10.0\pm 0.8\, (0.08)$\\
&
$N\pm \sqrt{N}$$(1/\sqrt{N})$&
$21\pm 5\, (0.22)$&
$8010\pm 90\, (0.01)$\\
\hline
\end{tabular}
\end{table}

To extract the single-spin cross section $d\Delta _{L}\sigma /d\xi $
from the single-spin asymmetry $A_{L}(\xi )$, the spin-averaged cross
sections $d\sigma /d\xi $ must be known well. Obviously, the uncertainty
in the knowledge of $d\sigma /d\xi $ must be small enough as compared
to the targeted uncertainty in the measurement of $d\Delta _{L}\sigma /d\xi $.
Therefore, an interesting question is: how well do we currently know
the unpolarized cross sections at RHIC energies, and how important
is their measurement in the upcoming RHIC experiments? The goal of
this section is to argue that at present the uncertainties in the
unpolarized cross sections at RHIC are sizeable; to reliably predict
the unpolarized rate, these uncertainties must be reduced by dedicated
RHIC measurements. 

Table~\ref{table:xsect} lists the unpolarized cross sections and
event rates for the processes $pp\rightarrow (W^{\pm }\rightarrow \ell \nu _{\ell })X$
and $pp\rightarrow (Z^{0}\rightarrow \ell \bar{\ell })X$ at RHIC
for various values of center-of-mass energies $\sqrt{s}$ and integrated
luminosities ${\mathcal{L}}$. The cross sections were derived at
NLO using the updated CTEQ6 parton distribution functions \cite{Pumplin:2002vw,Stump:2003yu}.
The cross sections $\sigma $ are calculated using the best-fit PDF
parametrization CTEQ6M. The uncertainty $\delta \sigma _{PDF}$ in
the cross section $\sigma $ is generated by the uncertainty in the
unpolarized parton luminosities, which are themselves phenomenological
functions known with a finite precision. This uncertainty is evaluated
with the help of the Hessian matrix analysis employed in Ref.~\cite{Pumplin:2002vw}.
Specifically, $\delta \sigma _{PDF}$ corresponds to the maximal variation
of $\sigma $ for all possible PDF parametrizations lying in the acceptable
range of $\chi ^{2}$. According to Table~\ref{table:xsect}, the
relative PDF uncertainties are of order $25\%$ at $\sqrt{s}=200$
GeV and $7-10\%$ at $\sqrt{s}=500$ GeV. These are the errors that
exist prior to the measurements at RHIC. 

The above PDF uncertainties are large because $W$ and $Z$ boson
production at the $pp$ collider RHIC probes the sea quark distributions
(primarily $\bar{u}(x)$ and $\bar{d}(x)$) at $x\gtrsim 0.1$. For
instance, the major part of $W$ boson events occurs in the central-rapidity
region $y_{W}\approx 0$ and corresponds to the Born-level momentum
fractions $x=Q/\sqrt{s}$ close to $0.4$ and $0.16$ at $\sqrt{s}=200$
GeV and $\sqrt{s}=500$ GeV, respectively. At such $x$ the sea quark
PDFs are not as constrained by the existing (mostly DIS) data, which
better determine the valence-dominated distributions $u(x)$ and $d(x).$
Therefore, RHIC provides the information that is complementary to
the DIS experiments at HERA and collider experiments at the $p\bar{p}$
collider Tevatron. 

\begin{figure}[p]
\begin{center}\includegraphics[  width=0.90\textwidth,
  keepaspectratio]{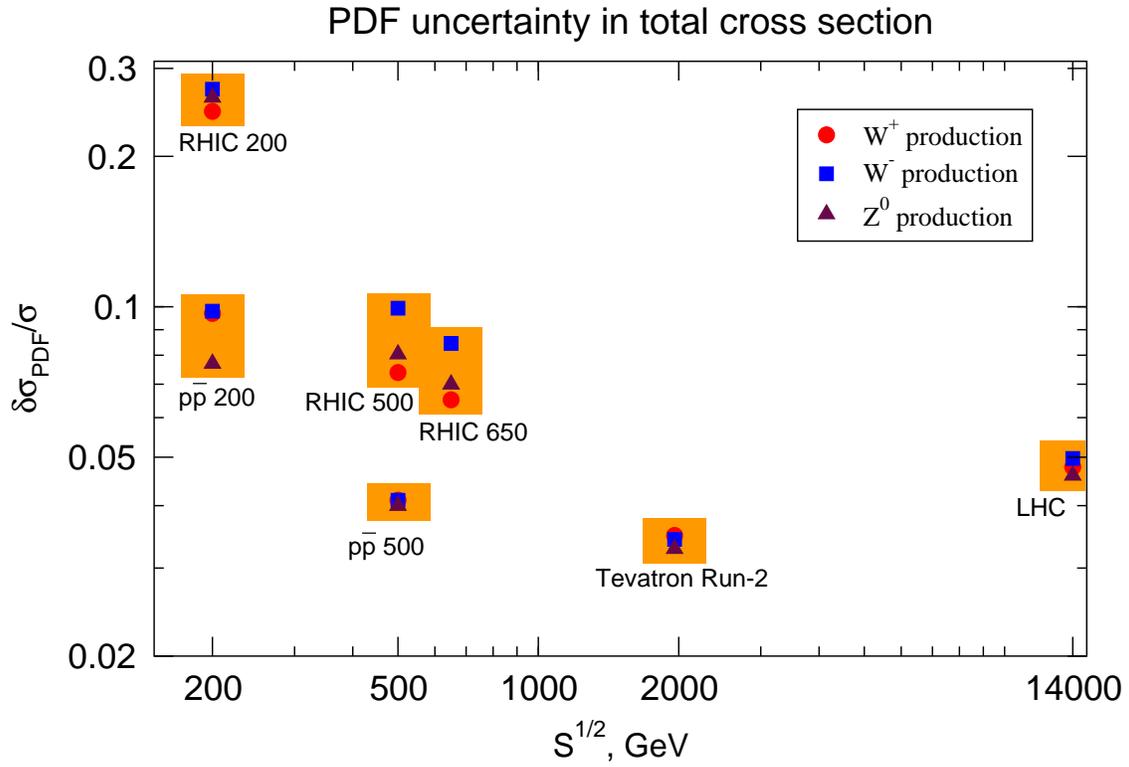}\end{center}

\caption{\label{fig:dsig_unpol} Uncertainties due to the PDF errors in the
unpolarized cross sections of $W$ and $Z^{0}$ boson production at
various colliders. The shown uncertainties are for RHIC at $\sqrt{s}=200,$
$500,$ and $650$ GeV; Tevatron Run-2; LHC; and fictitious $p\bar{p}$
colliders at $\sqrt{s}=200$ and $500$ GeV.}
\end{figure}

This point is illustrated by Fig.~\ref{fig:dsig_unpol}, which shows
the relative PDF uncertainty $\delta \sigma _{PDF}/\sigma $ in $W^{\pm }$
and $Z^{0}$ boson production at various existing colliders. To compare
the potential of $pp$ and $p\bar{p}$ colliders of the same center-of-mass
energy, this figure also shows the PDF uncertainties at fictitious
$p\bar{p}$ colliders with $\sqrt{s}=200$ and $500$ GeV. According
to the figure, the PDF uncertainties at RHIC substantially exceed
the uncertainties at the Tevatron and LHC. They are also up to 2.5
times larger than the PDF uncertainties at the fictitious $p\bar{p}$
colliders of RHIC energies. The PDF uncertainty is the largest in
$W^{-}$ boson production at RHIC, which is sensitive primarily to
the $d\bar{u}$ combination of the parton distributions.

On the other hand, for the planned integrated luminosities, RHIC will
measure the spin-averaged rates with better accuracy, particularly
in $W^{+}$ boson production. At $\sqrt{s}=500$ GeV and ${\mathcal{L}}=800\mbox {\, pb}^{-1}$,
the $2\sigma $ relative statistical errors ($2N^{-1/2}$) in the
rate are $0.6\%,$ $1.2\%$, and $2\%$ in $W^{+}$, $W^{-}$, and
$Z^{0}$ boson production, respectively. At $\sqrt{s}=200$ GeV and
${\mathcal{L}}=320\mbox {\, pb}^{-1}$, the respective errors are
$10\%$, $18\%,$ and $44\%$. We see that the projected statistical
errors in $W$ boson production are smaller than the current PDF errors.
Therefore, we conclude that the measurement of the spin-averaged cross
sections at RHIC will drastically reduce the uncertainty in the unpolarized
parton distributions in the probed region of $x$. We emphasize that
the measurement of the spin-averaged cross sections should be an important
part of the RHIC physics program, since it complements information
about the unpolarized PDFs from the other colliders and reduces systematic
uncertainties in the measured polarized cross sections.

\section{Reconstruction of spin asymmetries from RHIC data\label{sec:Reconstruction-of-spin}}

\subsection{Impact of multiple parton radiation on differential distributions}

To extract useful information from the raw data in $W$ boson production,
it is necessary to unfold the kinematical acceptance of the signal
events, which can only be reliably predicted by a resummation calculation,
similar to what has been done for the $W$ boson physics at the Tevatron
\cite{CPCsaba}. The correct lepton-level calculation is even more
indispensable given that RHIC cannot register the transverse energy
carried by the decay neutrino. 

\begin{figure}[p]
\begin{center}\includegraphics[  width=0.50\textwidth,
  keepaspectratio]{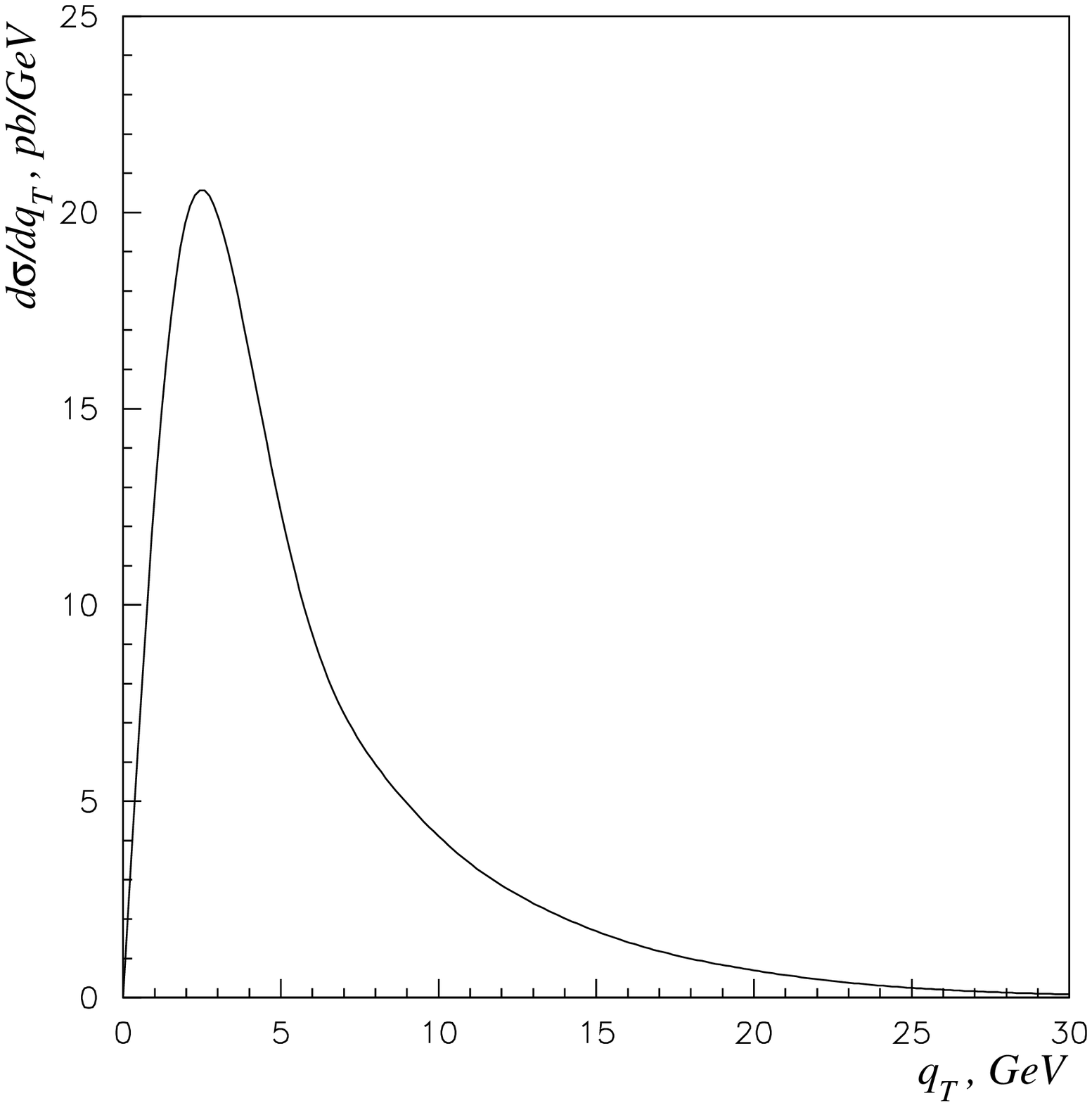}\includegraphics[  width=0.50\textwidth,
  keepaspectratio]{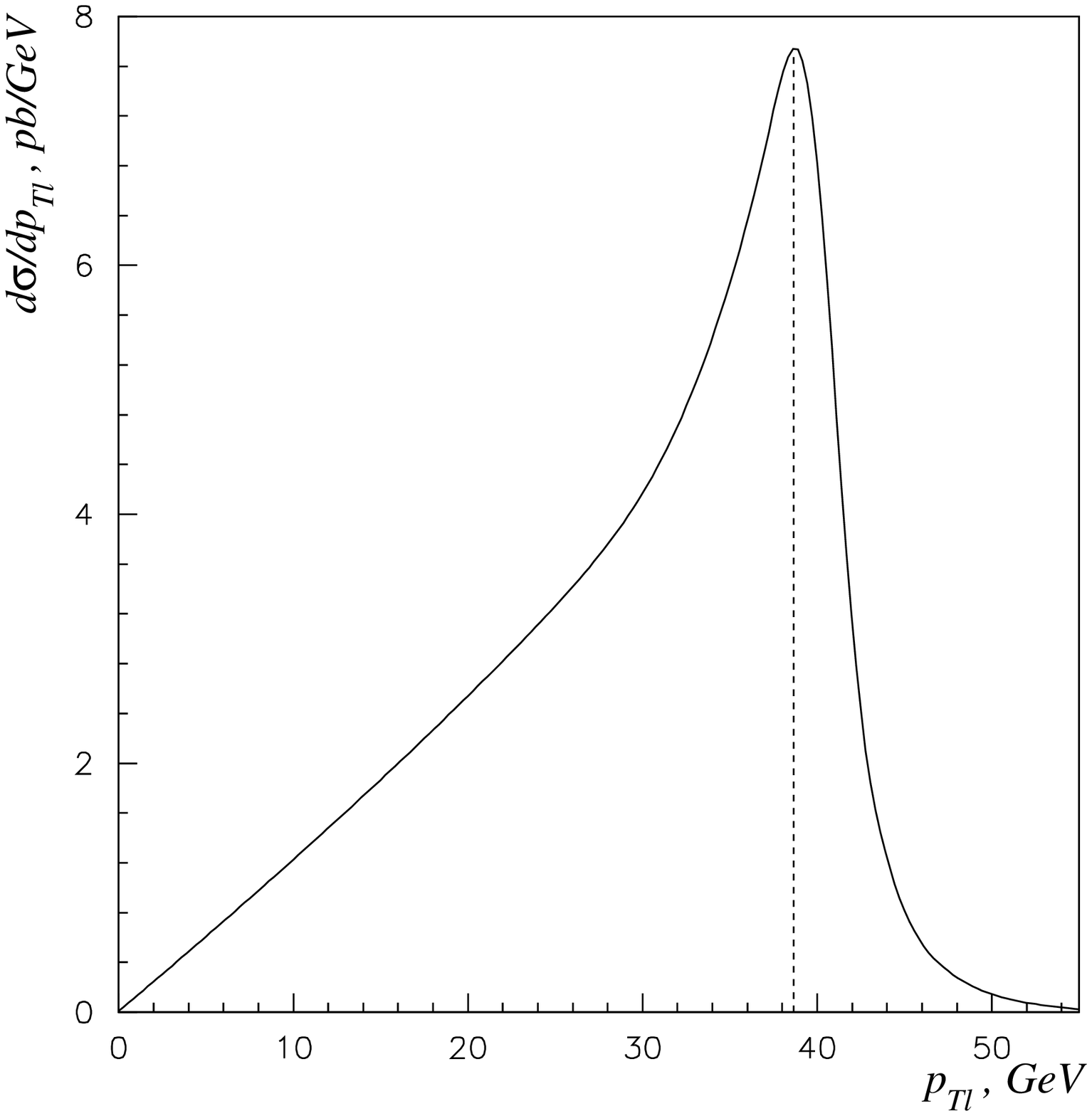}\\
(a)\hspace{2.9in}(b)\end{center}

\caption{\label{fig:qT} The unpolarized cross section for the process $pp\rightarrow (W^{+}\rightarrow \ell ^{+}\nu _{\ell })X$
at $\sqrt{s}=500$ GeV as a function of (a) the transverse momentum
$q_{T}$ of the $W^{+}$ boson and (b) the transverse momentum $p_{T\ell }$
of the charged lepton, as predicted by the resummation calculation.
CTEQ5M PDFs \cite{CTEQ5} were used.}
\end{figure}

For definiteness, let us concentrate on the production of $W^{+}$
bosons. Of course, all conclusions in this section hold for $W^{-}$
boson production as well. Fig.~\ref{fig:qT} shows the transverse
momentum distributions of the $W^{+}$ boson and charged lepton from
its decay, as predicted by the resummation package \textsc{Legacy-RhicBos}.
Here and after, the numerical results correspond to the c.m.~energy
$\sqrt{s}=500$ GeV and integrated luminosity ${\mathcal{L}}=800\mbox {\, pb}^{-1}$.
It is evident that most of the $W$ bosons are produced with small,
but non-zero transverse momenta. Such non-zero $q_{T}$ is acquired
through radiation of soft and collinear partons, which cannot be approximated
by finite-order perturbative calculations. In order to obtain reliable
predictions for differential cross sections, dominant logarithmic
terms $\alpha _{S}^{n}\ln ^{m}\left(q_{T}^{2}/Q^{2}\right)$ (where
$0\leq m\leq 2n-1$) associated with such radiation should be summed
through all orders of the perturbative series. 

\begin{figure}[p]
\subfigure[$pp \rightarrow (W^+ \rightarrow \ell^+ \nu_\ell) X$]{\includegraphics[  clip,
  width=0.50\textwidth,
  keepaspectratio]{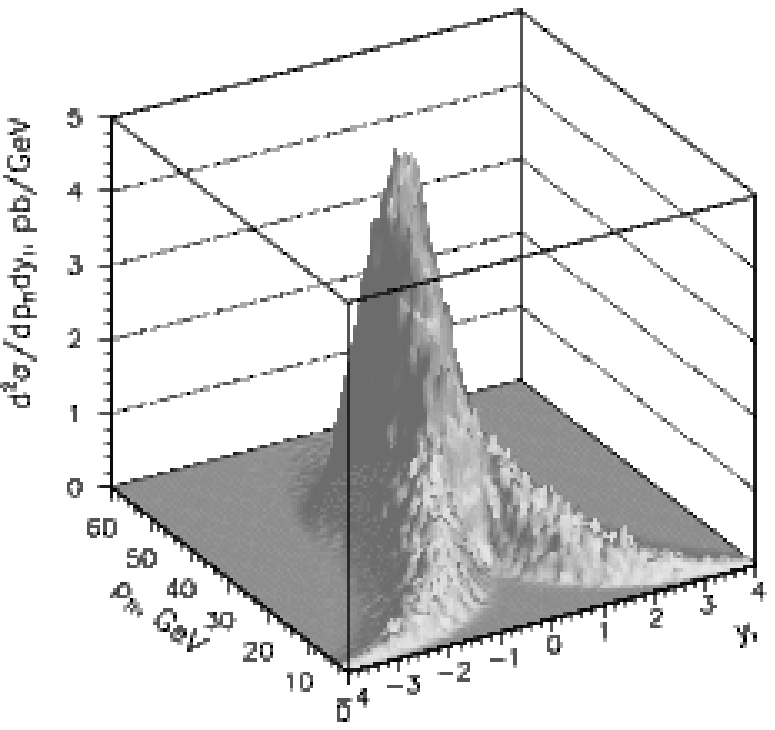}}\subfigure[$pp \rightarrow (W^- \rightarrow \ell^- \bar \nu_\ell) X$]{\includegraphics[  clip,
  width=0.50\textwidth,
  keepaspectratio]{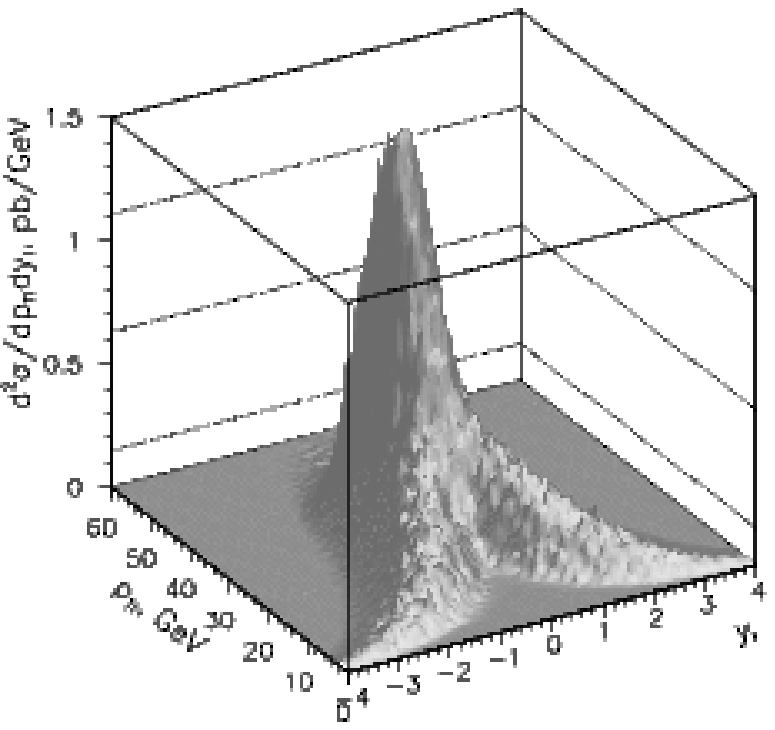}}

\subfigure[$\Delta_L p p \rightarrow (W^+ \rightarrow \ell^+ \nu_\ell) X$]{\includegraphics[  clip,
  width=0.50\textwidth,
  keepaspectratio]{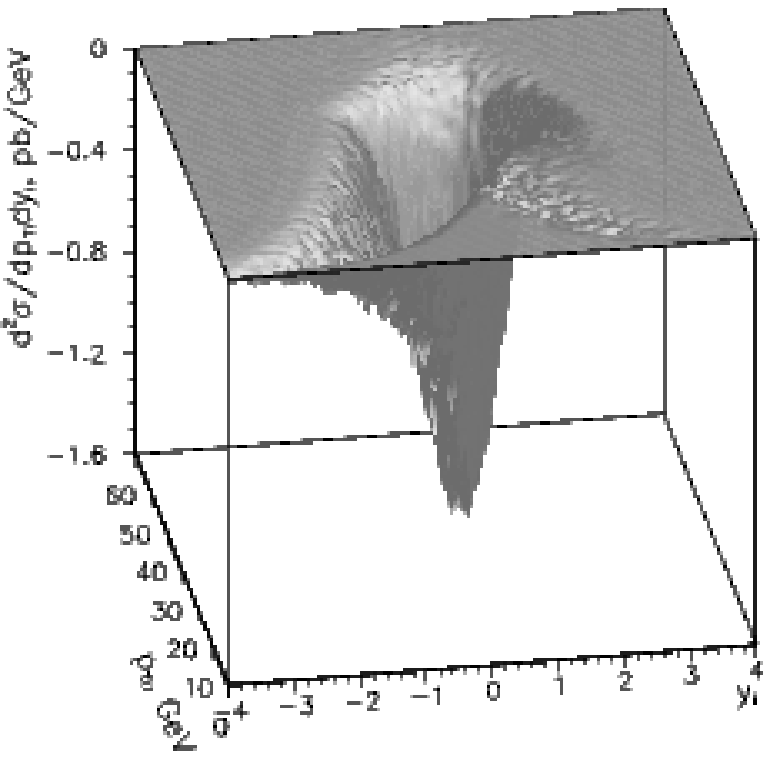}}\subfigure[$\Delta_L p p \rightarrow (W^- \rightarrow \ell^- \bar \nu_\ell) X$]{\includegraphics[  clip,
  width=0.50\textwidth,
  keepaspectratio]{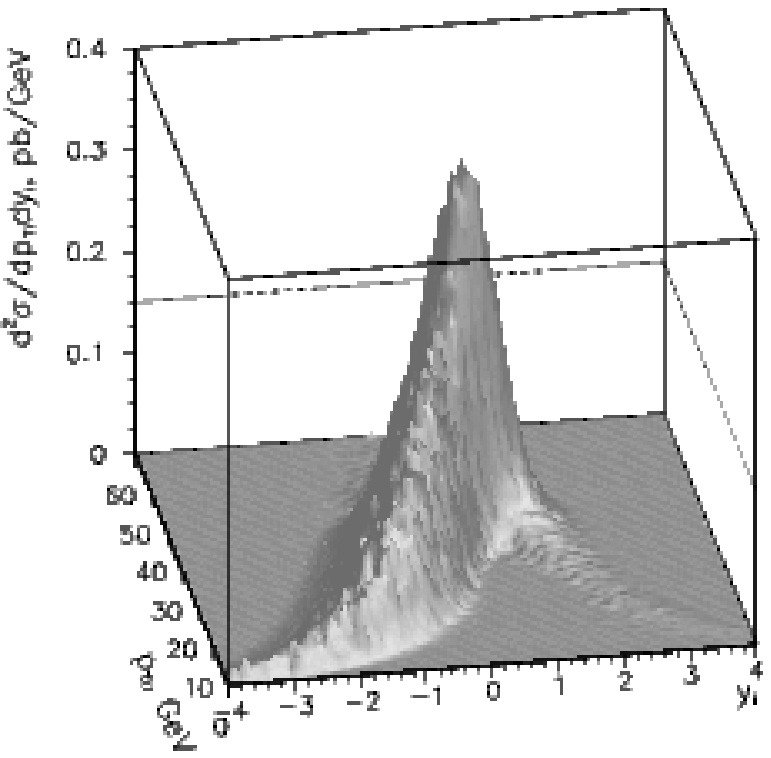}}

\caption{\label{fig:dsigma/dpTldyl}Resummed two-variable distributions $d^{2}\sigma /dp_{T\ell }dy_{\ell }$
for (a), (b) unpolarized and (c), (d) single-spin cross sections in
$W^{+}$ and $W^{-}$ boson production. The unpolarized and single-spin
cross sections are calculated using CTEQ5M \cite{CTEQ5} and GRSV-2000
\cite{Gluck:2000dy} standard PDF sets, respectively. A high-resolution version
of this figure is included in the online copy of the preprint available 
at http://hep.pa.msu.edu/\~{}nadolsky/RhicBos/home.html\#RhicBos\_references.}
\end{figure}

Furthermore, consider the distribution $d\sigma /dp_{T\ell }$ in
Fig.~\ref{fig:qT}(b). If the transverse momentum $q_{T}$ and width
$\Gamma _{W}$ of the $W$ boson were negligible (as in the Born-level
calculation), $d\sigma /dp_{T\ell }$ would show a singular Jacobian
peak exactly at $p_{T\ell }=M_{W}/2$. The smearing of the Jacobian
peak in Fig.~\ref{fig:qT}(b) is due to the multiple soft gluon radiation
and non-zero width of the $W$ boson. %
 The example of $d\sigma /dp_{T\ell }$ illustrates that multiple
parton radiation and heavy boson decay may drastically alter the Born-level
predictions. The resummation effects are important if the distribution
is sensitive to $q_{T}$ directly (as $d\sigma /dp_{T\ell }$), or
if the phase space is constrained in a region with slower convergence
of perturbation theory (as $d\sigma /dy_{\ell }$ at large $|y_{\ell }|$).
In the remainder of this section, we utilize this accurate calculation
to evaluate feasibility of the measurement of the boson-level asymmetry
$A_{L}(y_{W})$, which is commonly discussed in literature as the
most convenient observable to probe polarized sea quark distributions.
Many aspects of our discussion are based on the analysis of two-variable
distributions $d^{2}\sigma /(dp_{T\ell }dy_{\ell })$ in the transverse
momentum and rapidity of the observed charged lepton. These distributions
for the unpolarized and single-spin $W^{+}$ and $W^{-}$ boson production
are shown in Fig.~\ref{fig:dsigma/dpTldyl}. 

\subsection{$W$ boson rapidity at the Born level}

Consider the Born-level expression for the following asymmetry in
$W^{+}$ boson production induced by the up quark and down antiquark
via $u{\bar{d}}\rightarrow W^{+}\rightarrow \ell ^{+}\nu _{\ell }$:\begin{eqnarray}
A_{L}(y_{W},\theta ^{*}) & \equiv  & \frac{d\Delta _{L}\sigma /(dQ^{2}dy_{W}d\cos \theta ^{*})}{d\sigma /(dQ^{2}dy_{W}d\cos \theta ^{*})}\nonumber \\
 & = & \frac{-\Delta u(x_{A})\bar{d}(x_{B})(1+\cos \theta ^{*})^{2}+\Delta \bar{d}(x_{A})u(x_{B})(1-\cos \theta ^{*})^{2}}{u(x_{A})\bar{d}(x_{B})(1+\cos \theta ^{*})^{2}+\bar{d}(x_{A})u(x_{B})(1-\cos \theta ^{*})^{2}}.\label{ALyWtheta}
\end{eqnarray}
 Here $x_{A,B}\equiv (Q/\sqrt{s})e^{\pm y_{W}},$ and $\theta ^{*}$
is the polar angle of the charged lepton in the rest frame of the
$W$ boson, with the $z$ axis pointing in the moving direction of
the polarized proton beam. The single-spin asymmetry $A_{L}(y_{W})$
in the rapidity of the $W$ boson can be obtained from Eq.~(\ref{ALyWtheta})
by integrating $\cos \theta ^{*}$ out, i.e., integrating over the
moving direction of $\ell ^{+}$ and $\nu _{\ell }$: \begin{eqnarray}
A_{L}(y_{W}) & = & \frac{\int _{-1}^{1}\frac{d\Delta _{L}\sigma }{dQ^{2}dy_{W}d\cos \theta ^{*}}d\cos \theta ^{*}}{\int _{-1}^{1}\frac{d\sigma }{dQ^{2}dy_{W}d\cos \theta ^{*}}d\cos \theta ^{*}}=\frac{-\Delta u(x_{A})\bar{d}(x_{B})+\Delta \bar{d}(x_{A})u(x_{B})}{u(x_{A})\bar{d}(x_{B})+\bar{d}(x_{A})u(x_{B})}.\label{ALyW}
\end{eqnarray}
The resulting expression for $A_{L}(y_{W})$ is commonly discussed
in literature as an example of a simple observable, which is straightforwardly
sensitive to $\Delta u(x)$ and $\Delta \bar{d}(x).$ Similarly, the
Born-level asymmetry $A_{L}(y_{W})$ in $W^{-}$ boson production
is straightforwardly sensitive to $\Delta \bar{u}(x)$ and $\Delta d(x)$.
It has been also argued in Ref.~\cite{Gehrmann2} that ${\mathcal{O}}(\alpha _{S})$
corrections cancel to a good degree between the numerator and denominator
of $A_{L}(y_{W}).$ Unfortunately, as we will now demonstrate, the
measurement of $A_{L}(y_{W})$ is obstructed by the effects of $W$
boson decay and the limited detector acceptance, which seriously complicate
both the reconstruction of $y_{W}$ and determination of $A_{L}(y_{W})$
from the observed data. 

Let us first discuss the determination of the rapidity of the $W$
boson. Given that neither of the two RHIC detectors measures the missing
energy, the four-momentum of the $W$ boson cannot be deduced from
the momenta of its decay products. Therefore the rapidity of the $W$
boson cannot be measured directly. 

Despite the impossibility to measure $y_{W}$ in general, it has been
proposed \cite{Bland:1999gb,RHICOverview} to statistically \textit{choose}
the correct $y_{W}$ \textit{}\textit{\emph{in a certain kinematical
region}} based on several assumptions about the QCD dynamics of the
process. Let us outline the main idea of this method. 

Denote the rapidity and transverse momentum of the lepton in the lab
frame as $y_{\ell }$ and $p_{T\ell }$, respectively. Similarly,
denote the corresponding variables in the rest frame of the $W$ boson
as $y_{\ell }^{\prime }$ and $p_{T\ell }^{\prime }$, respectively.
In the naive Born approximation, the $W$ bosons have zero width ($Q=M_{W}$)
and zero transverse momentum ($q_{T}=0$). Therefore, the lepton variables
in the different reference frames are related as\begin{eqnarray}
y_{\ell } & = & y_{\ell }^{\prime }+y_{W},\label{yl}\\
p_{T\ell } & = & q_{T}+p_{T\ell }^{\prime }\begin{array}[t]{c}
 =\\
 \stackrel{\uparrow }{q_{T}=0}\end{array}
p_{T\ell }^{\prime },\label{pTl}
\end{eqnarray}
 where $y_{\ell }^{\prime }$ and $p_{T\ell }^{\prime }$ are the
functions of the polar angle $\theta ^{*}$ of the lepton in the $W$
boson rest frame:\begin{eqnarray}
y_{\ell }^{\prime } & = & \frac{1}{2}\ln \frac{1+\cos \theta ^{*}}{1-\cos \theta ^{*}},\label{ylprime}\\
p_{T\ell }^{\prime } & = & \frac{Q}{2}\sin \theta ^{*}\begin{array}[t]{c}
 \begin{array}[t]{c}
 =\\
 \stackrel{\uparrow }{Q=M_{W}}\end{array}
\end{array}
\frac{M_{W}}{2}\sin \theta ^{*}.\label{pTlprime}
\end{eqnarray}
The comments below the equations show the assumptions involved in
obtaining the rightmost expressions in Eqs\@.~(\ref{pTl}) and (\ref{pTlprime}).%
\begin{figure}[p]
\begin{center}\includegraphics[  width=0.50\textwidth,
  keepaspectratio]{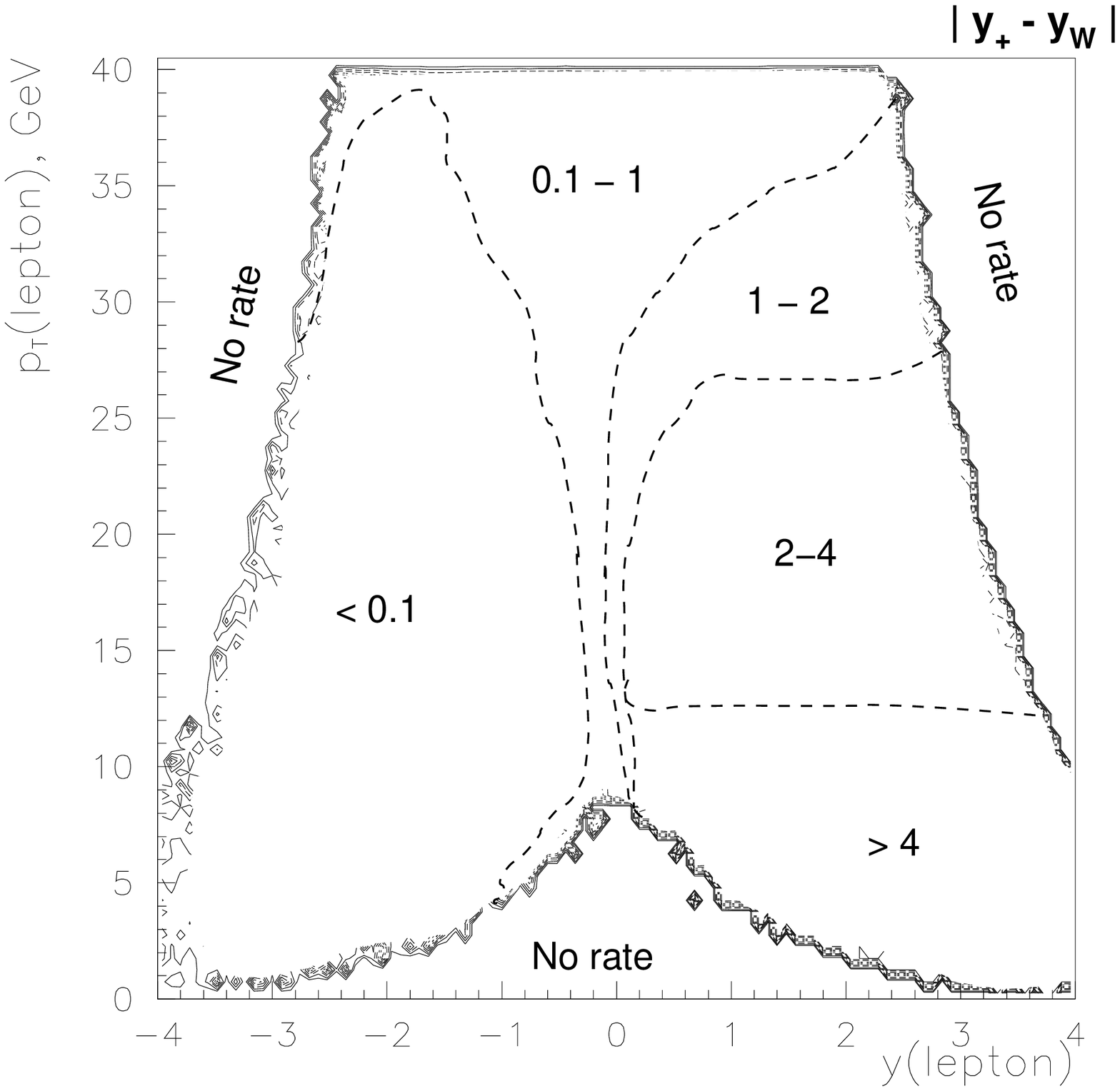}\includegraphics[  width=0.50\textwidth,
  keepaspectratio]{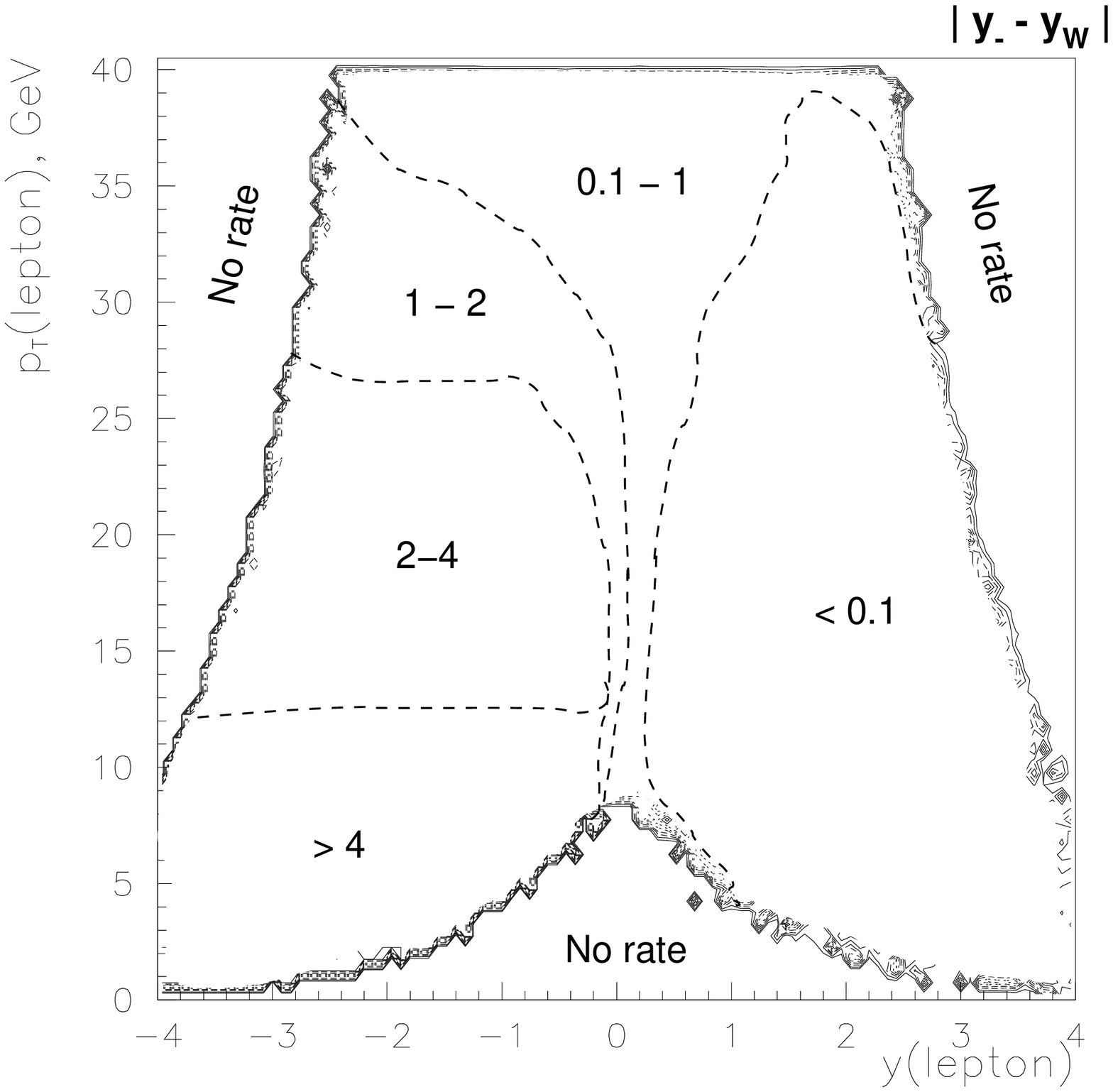}\\
(a)\hspace{2.9in}(b)\end{center}

\caption{\label{fig:butterflies} The average differences between the exact
value of $y_{W}$ and two approximate solutions $y_{\pm }$ of Eqs.~(\ref{yl})-(\ref{pTlprime})
as a function of the charged lepton's rapidity $y_{\ell }$ and transverse
momentum $p_{T\ell }.$ The resummation calculation and CTEQ5M PDFs
\cite{CTEQ5} were used. The numbers inside the {}``butterfly''
shapes show the magnitude of the differences $|y_{+}-y_{W}|$ and
$|y_{-}-y_{W}|$. According to the figure, the solutions $y_{+}$
and $y_{-}$ provide good approximations in the limits $y_{\ell }\ll 0$
and $y_{\ell }\gg 0,$ respectively. The dashed lines approximately
mark the regions of the shown rapidity differences. A more detailed
color version of the figure is available in the online copy of the preprint
at 
http://hep.pa.msu.edu/\textasciitilde{}nadolsky/RhicBos/home.html\#RhicBos\_references.
}
\end{figure}

For a given set of $y_{\ell }$ and $p_{T\ell }$, Eqs.~(\ref{yl})-(\ref{pTlprime})
imply two solutions $y_{+}$ and $y_{-}$ (corresponding to $\theta ^{*}$
and $\pi -\theta ^{*}$) for $y_{W}$. One of them can be correctly
chosen if the magnitude of the rapidity $y_{\ell }$ of the charged
lepton is large (i.e., if the lepton is observed in the forward or
backward rapidity region). In these special regions, one of the two
solutions for $y_{W}$ can always be discarded, because it takes a
value outside of the allowed rapidity range of the collider. In other
words, the chosen solution is the one with the smaller magnitude.
However, in reality, $q_{T}$ never vanishes, as predicted by the
resummation calculation (see Fig.~\ref{fig:qT}(a)). Similarly, $\Gamma _{W}$
is not zero. Therefore, the above approximation never holds exactly,
and the size of the error is determined by the difference between
the Born-level and exact dynamics. 

\begin{figure}[p]
\begin{center}\includegraphics[  clip,
  height=8cm,
  keepaspectratio]{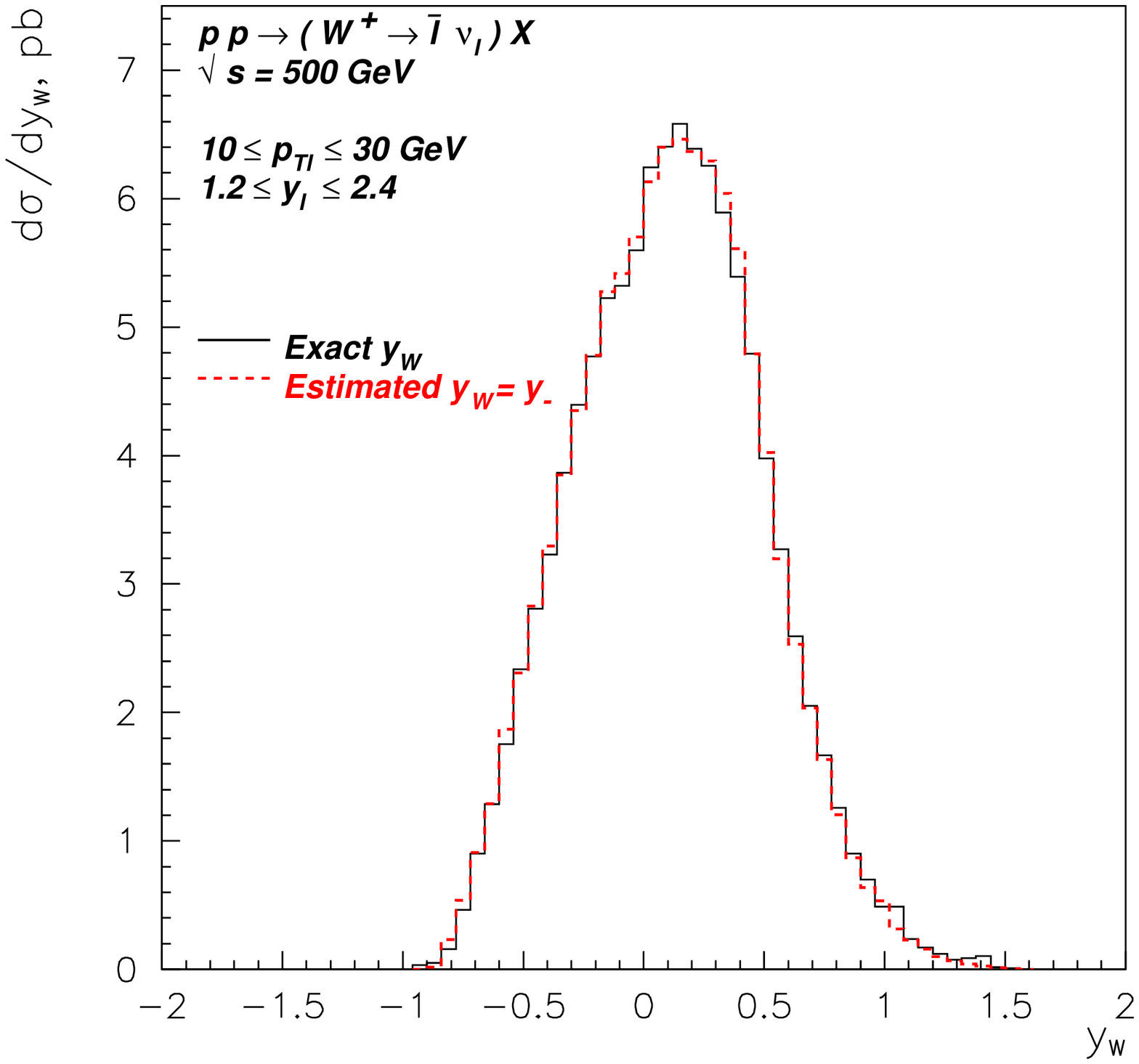}\end{center}

\caption{\label{fig:y_w_app}Comparison of the distributions $d\sigma /dy_{-}$
and $d\sigma /dy_{W}$ in the region of the validity of the approximate
solution $y_{-}$, as predicted by the resummation calculation. The
constraints $1.2\leq y_{\ell }\leq 2.4$ and $p_{T\ell }\geq 10$
GeV imitate kinematical cuts for a measurement with the \PHENIX muon
detector. The cut $p_{T\ell }\leq 30$ GeV suppresses contributions
with large deviations of $y_{-}$ from $y_{W}$ (cf.~Fig.~\ref{fig:butterflies}(b)).
CTEQ5M PDFs \cite{CTEQ5} were used.}
\end{figure}

By comparing the leading order, NLO, and resummation calculations,
we have found that the approximation holds reasonably well when $q_{T}\ll M_{W}$,
$p_{T\ell }^{\prime }\approx p_{T\ell }$, and $y_{\ell }$ is large.
This point is illustrated in Fig.~\ref{fig:butterflies}, which shows
the statistically averaged differences $\langle (|y_{\pm }-y_{W}|)\sigma \rangle /\langle \sigma \rangle $
between the exact $y_{W}$ and each of the two solutions $y_{\pm }$
in the two-dimensional plane of $y_{\ell }$ and $p_{T\ell }$, predicted
by the resummation calculation. It can be seen that the difference
is small for one of the solutions in the respective region $y_{\ell }\gtrsim 1$
or $y_{\ell }\lesssim -1$, and $p_{T\ell }\lesssim 25-30$ GeV. At
the same time, the difference is unacceptably large for the other
solution. Correspondingly, in each special region the distribution
in the correct solution $y_{+}$ or $y_{-}$ closely reproduces the
distribution in true $y_{W}$. For example, Fig.~\ref{fig:y_w_app}
shows that the distributions $d\sigma /dy_{-}$ agrees well with the
distribution $d\sigma /dy_{W}$ in the region $10\mbox {\, GeV}\leq p_{T\ell }\leq 30\mbox {\, GeV}$
and $1.2\leq y_{\ell }\leq 2.4$. %

Except for the special regions in Fig.~\ref{fig:butterflies}, the
approximate solutions $y_{\pm }$ do not agree well with the exact
$y_{W}$. An important feature to note is that in the special regions
the event rate is small. As shown in Fig.~\ref{fig:dsigma/dpTldyl},
most of the event rate comes from $p_{T\ell }\sim M_{W}/2$ and $y_{\ell }\sim 0$.
Furthermore, the magnitude of the error in $y_{W}$ depends on the
shape of the cross section, which, of course, is unknown \emph{a priori}
in the polarized case.

\subsection{Distortion effects on $A_{L}(y_{W}$)}

Let us now turn to another problem in the measurement of $A_{L}(y_{W})$:
distortions in the deduced shape of $A_{L}(y_{W})$ due the experimental
cuts. To obtain $A_{L}(y_{W})$, the experimental cross sections have
to be integrated over all possible directions of the lepton motion,
i.e., over $\cos \theta ^{*}$ in the interval $-1\leq \cos \theta ^{*}\leq 1$
(see Eq.~(\ref{ALyW}) for an example of such integration in the
Born-level analysis). Unfortunately, the acceptance of RHIC detectors
does not permit such integration, so that the resulting asymmetry
has the residual dependence on the acceptance range in $\cos \theta _{*}$
and does not agree with the true asymmetry $A_{L}(y_{W})$. 

Indeed, $\cos \theta ^{*}$ is related to the rapidity of the lepton
$y_{\ell }$ via Eqs.~(\ref{yl}) and (\ref{ylprime}):\[
y_{\ell }=y_{W}+\frac{1}{2}\ln \frac{1+\cos \theta ^{*}}{1-\cos \theta ^{*}}.\]
 Therefore, the restrictions on the range of $y_{\ell }$ reduce the
range of integration over $\cos \theta ^{*}.$ Specifically, the \PHENIX
muon detectors are able to detect a muon if the muon rapidity $y_{\mu }$
and azimuthal angle $\phi _{\mu }$ satisfy $1.2\leq |y_{\mu }|\leq 2.4$
and $0\leq \phi _{\mu }\leq 2\pi $. The \PHENIX electromagnetic
calorimeter can register electrons with $|y_{e}|\leq 0.35$ and $0\leq \phi _{e}\leq \pi $.
The \STAR detector registers electrons in the barrel electromagnetic
calorimeter that covers $|y_{e}|\leq 1.0$ and $0\leq \phi _{e}\leq 2\pi $.
In addition, a selection cut $p_{T\ell }\gtrsim 10-20$ GeV is expected
to be imposed on the charged leptons to suppress background contributions,
which also constrains $\sin \theta ^{*}$ through Eqs.~(\ref{pTl})
and (\ref{pTlprime}). 

\begin{figure}[p]
\begin{center}\includegraphics[  bb=0bp 0bp 531bp 505bp,
  clip,
  height=8cm,
  keepaspectratio]{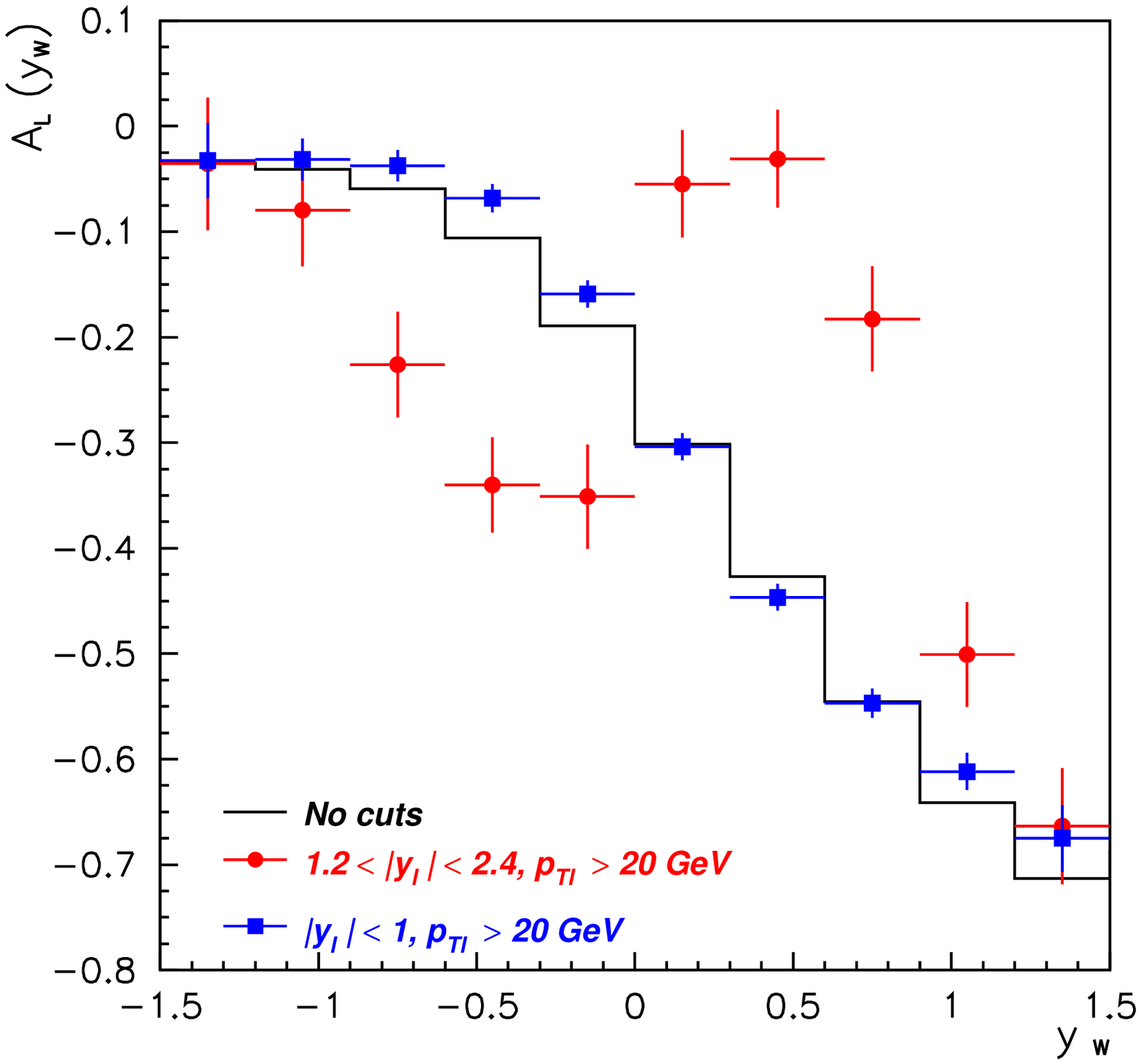}\end{center}

\caption{\label{fig:ALyvscuts}Dependence of the asymmetry $A_{L}(y_{W})$
on the cuts imposed on the momentum of the observed charged lepton
in the process $\Delta _{L}pp\rightarrow (W^{+}\rightarrow \ell ^{+}\nu _{\ell })X$,
as predicted by the resummation calculation. The GRSV-2000 standard
set \cite{Gluck:2000dy} of the polarized PDFs was used. The projected
statistical errors in $A_{L}(y_{W})$ (shown by error bars) are estimated
with the help of Eq.~(\ref{deltaAL}) for the bin sizes shown in
the figure. }
\end{figure}

The effect of the above cuts is demonstrated in Fig.~\ref{fig:ALyvscuts},
which shows the single-spin asymmetry in $W^{+}$ boson production
calculated without constraints on $y_{\ell }$ and $p_{T\ell }$ (solid
line), as well as with the constraints $1.2<|y_{\ell }|<2.4,$ $p_{T\ell }>20\mbox {\, GeV}$
(circles) and $|y_{\ell }|<1,$ $p_{T\ell }>20$ GeV (boxes). The
results shown here are derived using the resummation calculation.%
\footnote{For the asymmetries $A_{L}(y_{W})$, the resummation and NLO calculations
give close predictions (within one estimated statistical error $\delta A_{L}$),
even though these calculations predict quite different shapes for the
unpolarized and polarized differential cross sections. %
} According to the figure, there is a substantial difference between
the asymmetries calculated with and without experimental cuts. This
difference arises due to the different dependence of the unpolarized
and polarized cross sections on angular distributions of the charged
leptons, which affects the asymmetry because $\cos \theta ^{*}$ is
not integrated out completely. %

The findings in this section can be summarized as follows. The extraction
of $A_{L}(y_{W})$ from RHIC data is not an easy task, because (a)
the direct measurement of $y_{W}$ is impossible, and (b) the reconstructed
$A_{L}(y_{W})$ is distorted by the experimental acceptance cuts.
In particular, the two-fold ambiguity in the determination of $y_{W}$
is unavoidable at $y_{\ell }\sim 0$ and $p_{T\ell }\sim M_{W}/2,$
i.e., in the region with the largest cross section. To take the full
advantage of the large event rate at RHIC, we ought to seek an alternative
to $A_{L}(y_{W})$, which is free of the complications discussed above.
In the next section, we turn our attention to the spin asymmetries
of the lepton-level cross sections $d\sigma /dy_{\ell }$ and $d\sigma /dp_{T\ell }$,
which may serve as such alternative observables. Based on the resummation
calculation, we analyze the sensitivity of the lepton-level asymmetries
to the choice of the experimental cuts. We also present theoretical
predictions for various PDF sets.

\section{Single-spin asymmetries at the lepton level\label{sec:Lepton-level-asymmetries}}

\subsection{Impact of the cuts and soft radiation on lepton-level cross sections}

Let us first separately discuss the unpolarized and polarized cross
sections. To study the detector effects on the shape of various distributions,
we will compare the cross sections for the kinematical acceptance
of \PHENIX and \STAR detectors, as well as cross sections without
any kinematical cuts. We also compare the resummation calculation
to the NLO calculation, with the latter performed in the phase space
slicing method for the separation parameter $q_{T}^{sep}=1.6$ GeV.
We concentrate on the distributions $d\sigma /dp_{T\ell }$, where
the differences between the resummed and NLO predictions are the most
visible.%
\begin{figure}[p]
\begin{center}\subfigure[No cuts]{\includegraphics[  width=0.50\textwidth,
  keepaspectratio]{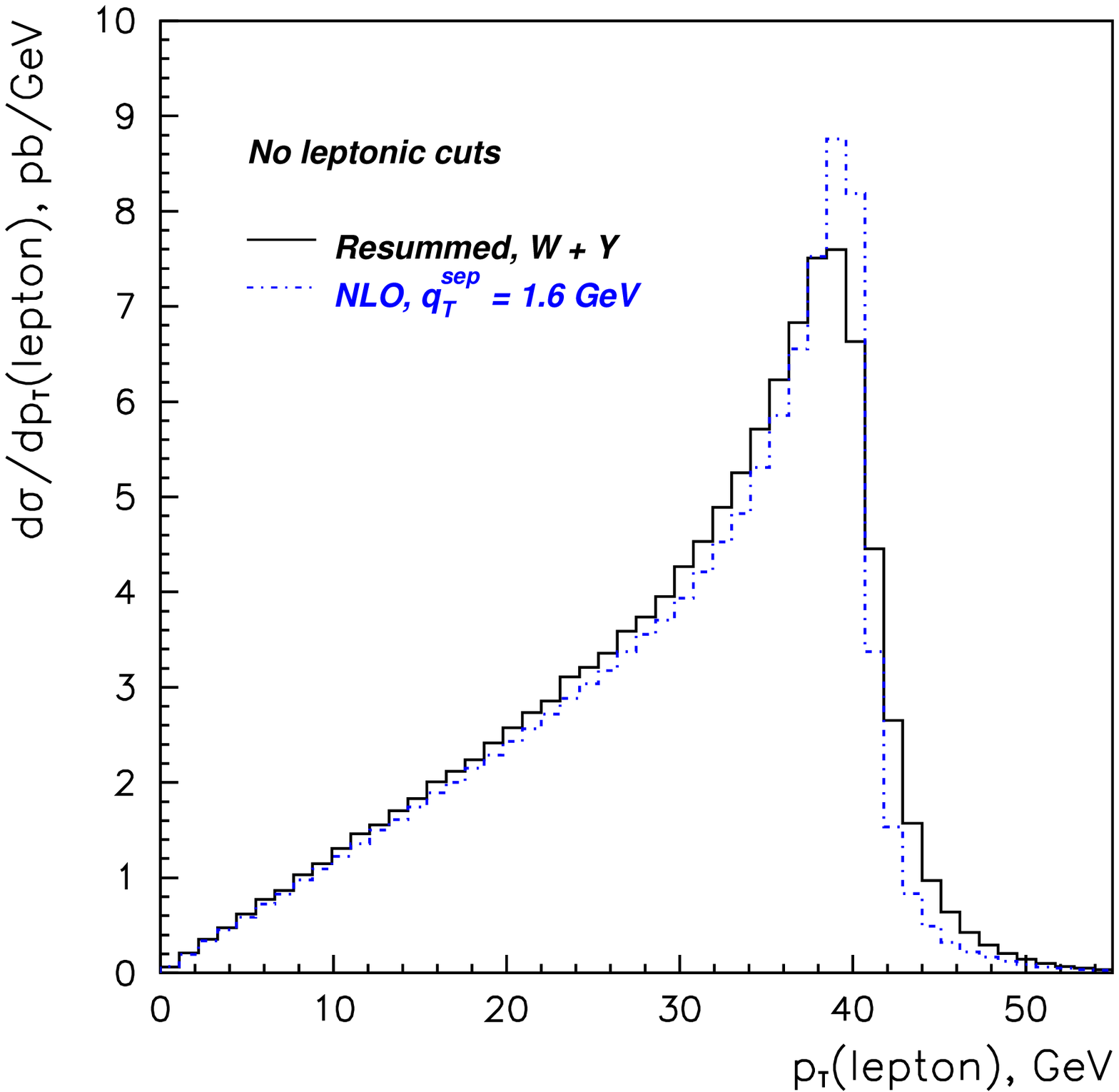}}\subfigure[$1.2\leq |y_\ell |\leq 2.4$]{\includegraphics[  width=0.50\textwidth,
  keepaspectratio]{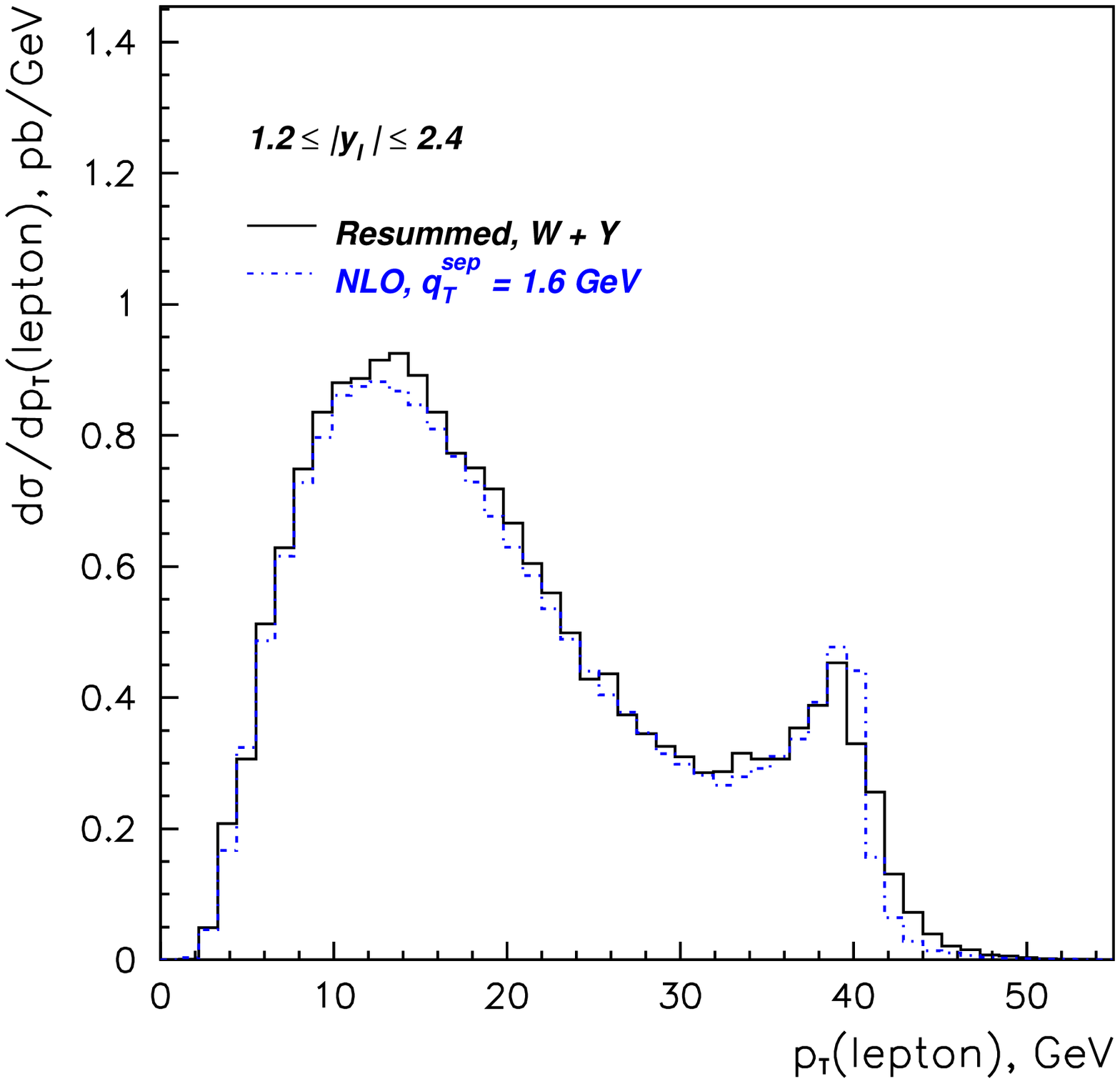}} \subfigure[$-1 \leq y_\ell \leq 1$]{\includegraphics[  width=0.50\textwidth,
  keepaspectratio]{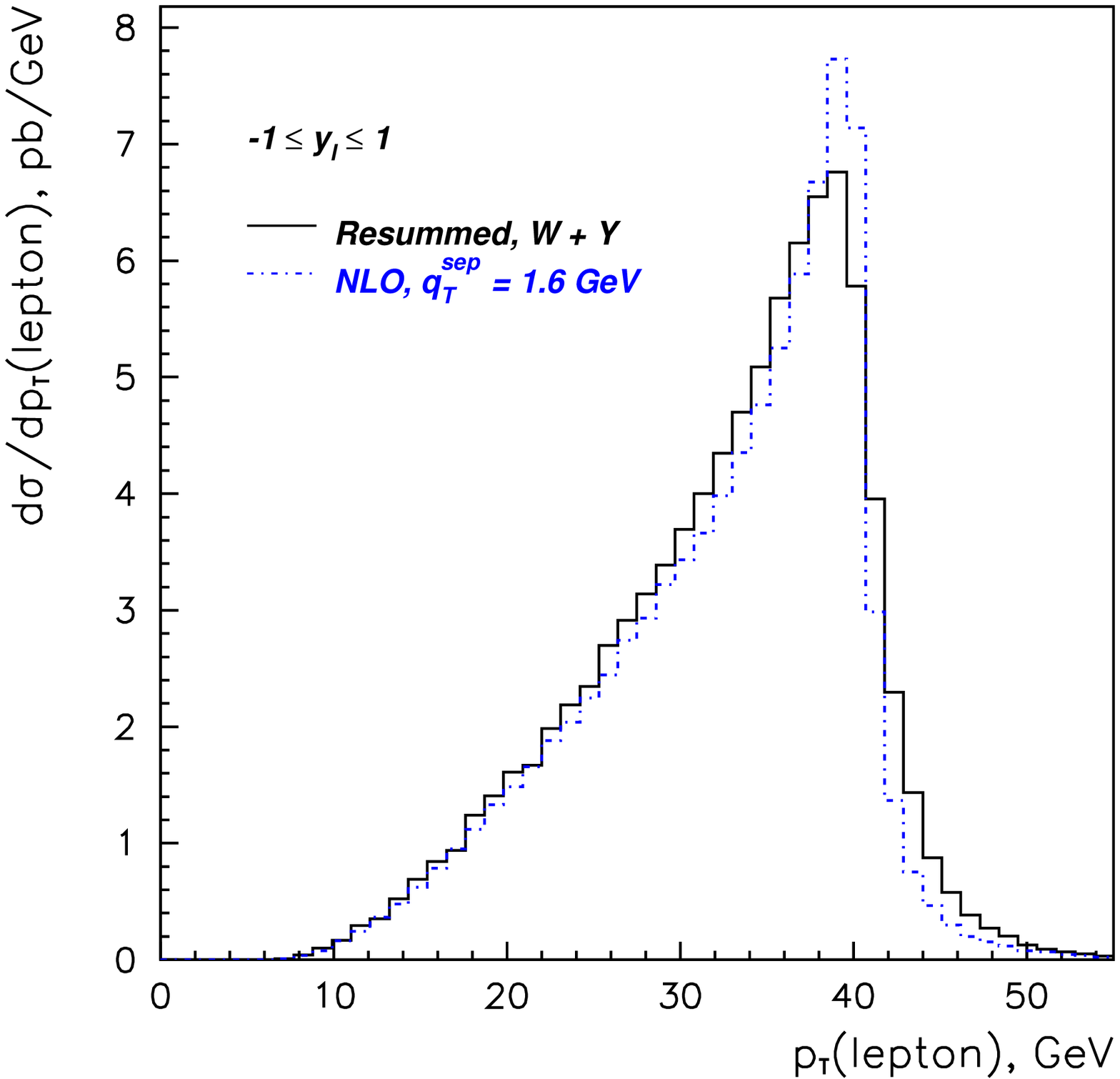}} \end{center}

\caption{\label{fig:ptl_unp_wp} Unpolarized resummed (solid) and NLO (dashed)
cross sections $d\sigma /dp_{T\ell }$ in the process $pp\rightarrow (W^{+}\rightarrow \ell ^{+}\nu _{\ell })X$
for various cuts on the lepton's rapidity. CTEQ5M PDFs \cite{CTEQ5}
were used.}
\end{figure}

Fig.~\ref{fig:ptl_unp_wp} presents the distribution $d\sigma /p_{T\ell }$
for the positively charged leptons from the decay of $W^{+}$ bosons.
The shape of the NLO cross sections (dashed curves) for $p_{T\ell }$
around and above $M_{W}/2$ (i.e., the Jacobian peak) is strongly
affected by the arbitrary parameter $q_{T}^{sep}$ employed in this
calculation. Hence, the only reliable predictions for this part of
the phase space are given by the resummed cross sections (shown as
the solid curves). For smaller $p_{T\ell }$, the resummation calculation
agrees with the NLO calculation, because it is formulated to match
the NLO calculation in the region where the multiple gluon emission
is not important. Furthermore, without any kinematic cuts, the integrated
rate of the resummation calculation agrees with that of the NLO calculation
to a percent level. As shown in Fig.~\ref{fig:ptl_unp_wp}(b), the
selection of the leptons in the large rapidity region distorts the
shape of the $p_{T\ell }$ distribution by suppressing the contributions
near the Jacobian peak. On the other hand, the selection of the central-rapidity
region $|y_{e}|<1.0$ (cf.~Fig.~\ref{fig:ptl_unp_wp}(c)) suppresses
the contributions with $p_{T\ell }\lesssim 10$ GeV. Both features
can be understood from the shape of the differential distribution
$d^{2}\sigma /(dp_{T\ell }dy_{\ell })$ given in Fig.~\ref{fig:dsigma/dpTldyl}(a).
According to this figure, the small (large) rapidity $|y_{\ell }|$
statistically corresponds to the large (small) transverse momentum
$p_{T\ell }$. 

\begin{figure}[p]
\begin{center}\subfigure[No cuts]{\includegraphics[  width=0.50\textwidth,
  keepaspectratio]{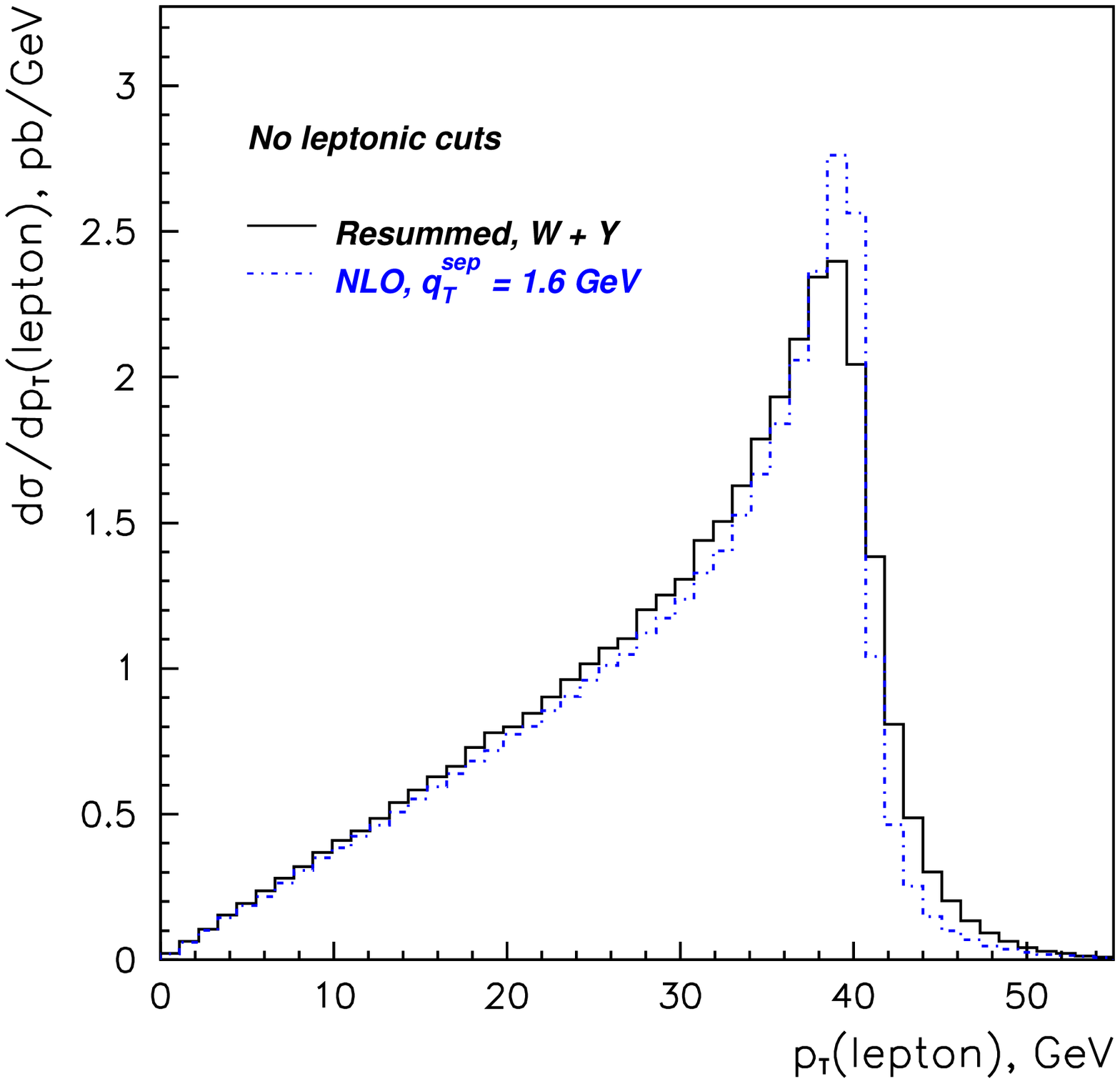}}\subfigure[$1.2\leq |y_\ell |\leq 2.4$]{\includegraphics[  width=0.50\textwidth,
  keepaspectratio]{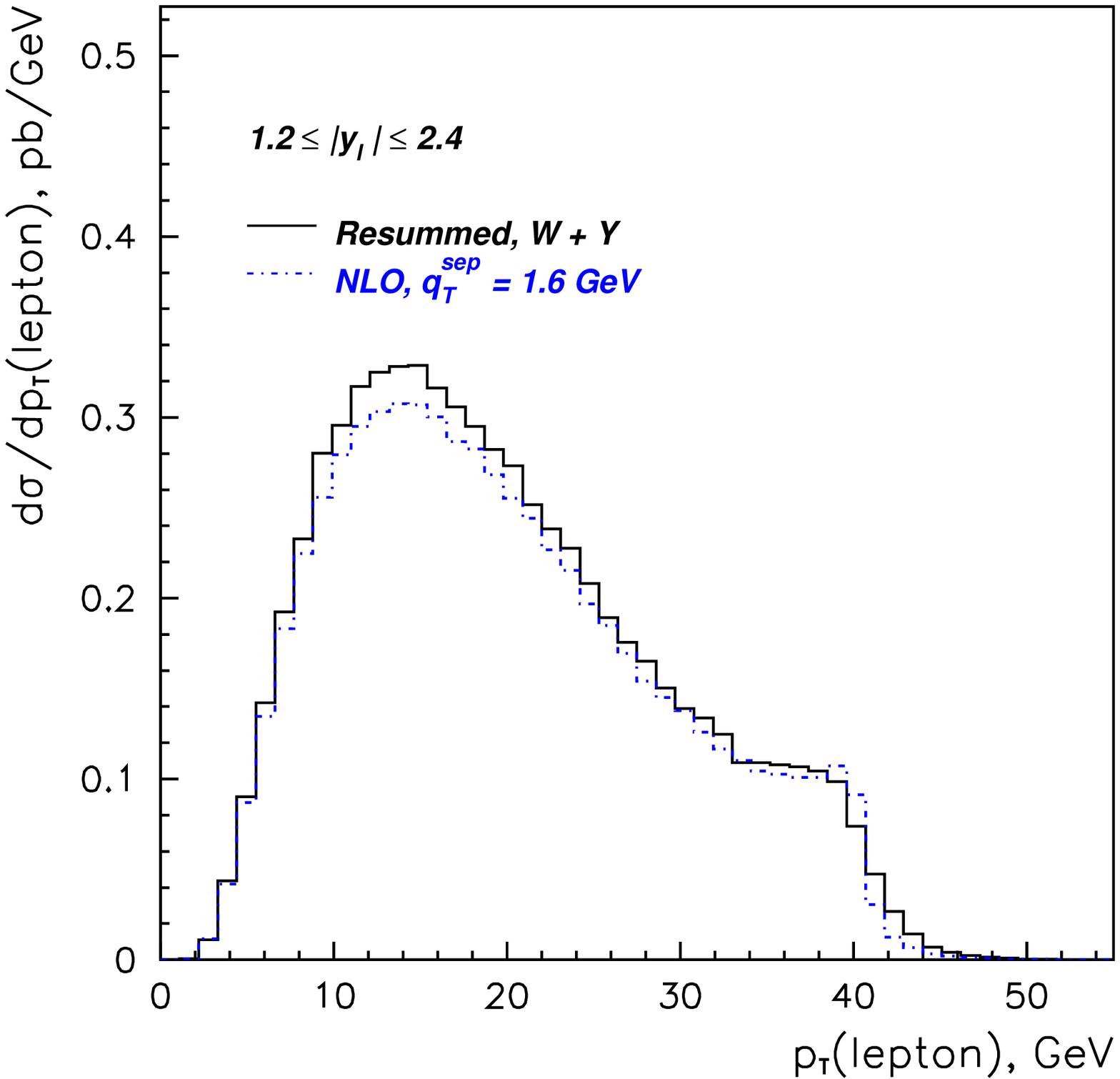}} \subfigure[$-1\leq y_\ell \leq 1$]{\includegraphics[  width=0.50\textwidth,
  keepaspectratio]{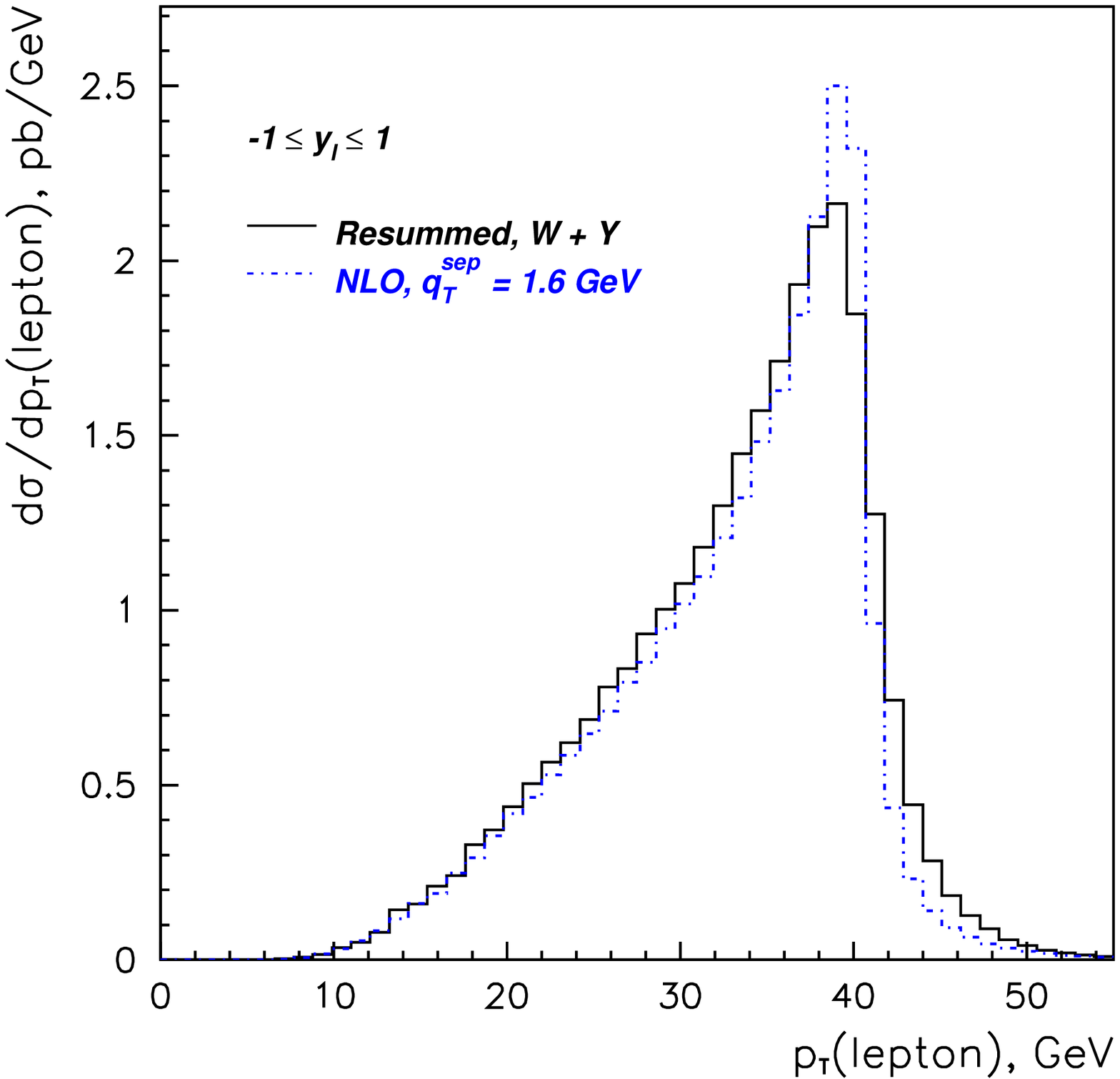}} \end{center}

\caption{\label{fig:ptl_unp_wm}Unpolarized resummed (solid) and NLO (dashed)
cross sections $d\sigma /dp_{T\ell }$ in the process $pp\rightarrow (W^{-}\rightarrow \ell ^{-}\bar{\nu }_{\ell })X$
for various cuts on the lepton's rapidity. CTEQ5M PDFs \cite{CTEQ5}
were used. }
\end{figure}

As a $pp$ collider, RHIC produces different numbers of $W^{-}$ and
$W^{+}$ bosons because of different parton luminosities.  Due to
the different rapidity distributions for $W^{+}$ and $W^{-}$ bosons,
the rapidity cuts have a different effect on $d\sigma /dp_{T\ell }$
in these processes. Fig.~\ref{fig:ptl_unp_wm} shows the distributions
$d\sigma /dp_{T\ell }$ in the process $pp\rightarrow (W^{-}\rightarrow \ell ^{-}\bar{\nu }_{\ell })X$.
We find that the shape of the distributions without cuts shown in
Fig.~\ref{fig:ptl_unp_wm}(a) is very close to the shape of the corresponding
distribution in Fig.~\ref{fig:ptl_unp_wp}(a), even though the overall
normalizations are obviously different. On the other hand, the shapes
do not coincide in the presence of the cuts, in particular in the
case of the \PHENIX selection cuts shown in Figs.~\ref{fig:ptl_unp_wp}(b)
and \ref{fig:ptl_unp_wm}(b). %

\begin{figure}[p]
\begin{center}\subfigure[No cuts]{\includegraphics[  width=0.50\textwidth,
  keepaspectratio]{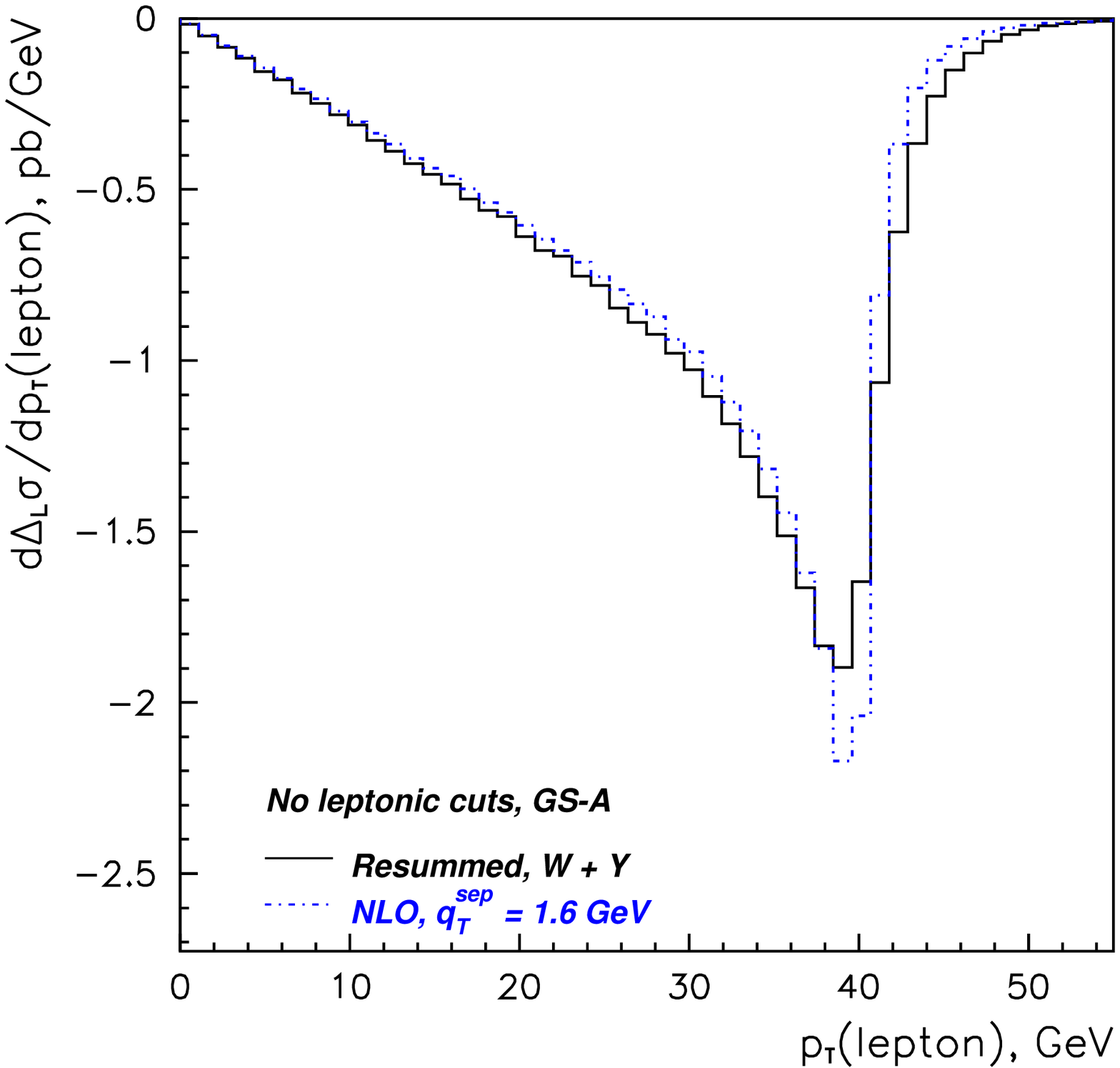}}\subfigure[$1.2\leq |y_\ell |\leq 2.4$]{\includegraphics[  width=0.50\textwidth,
  keepaspectratio]{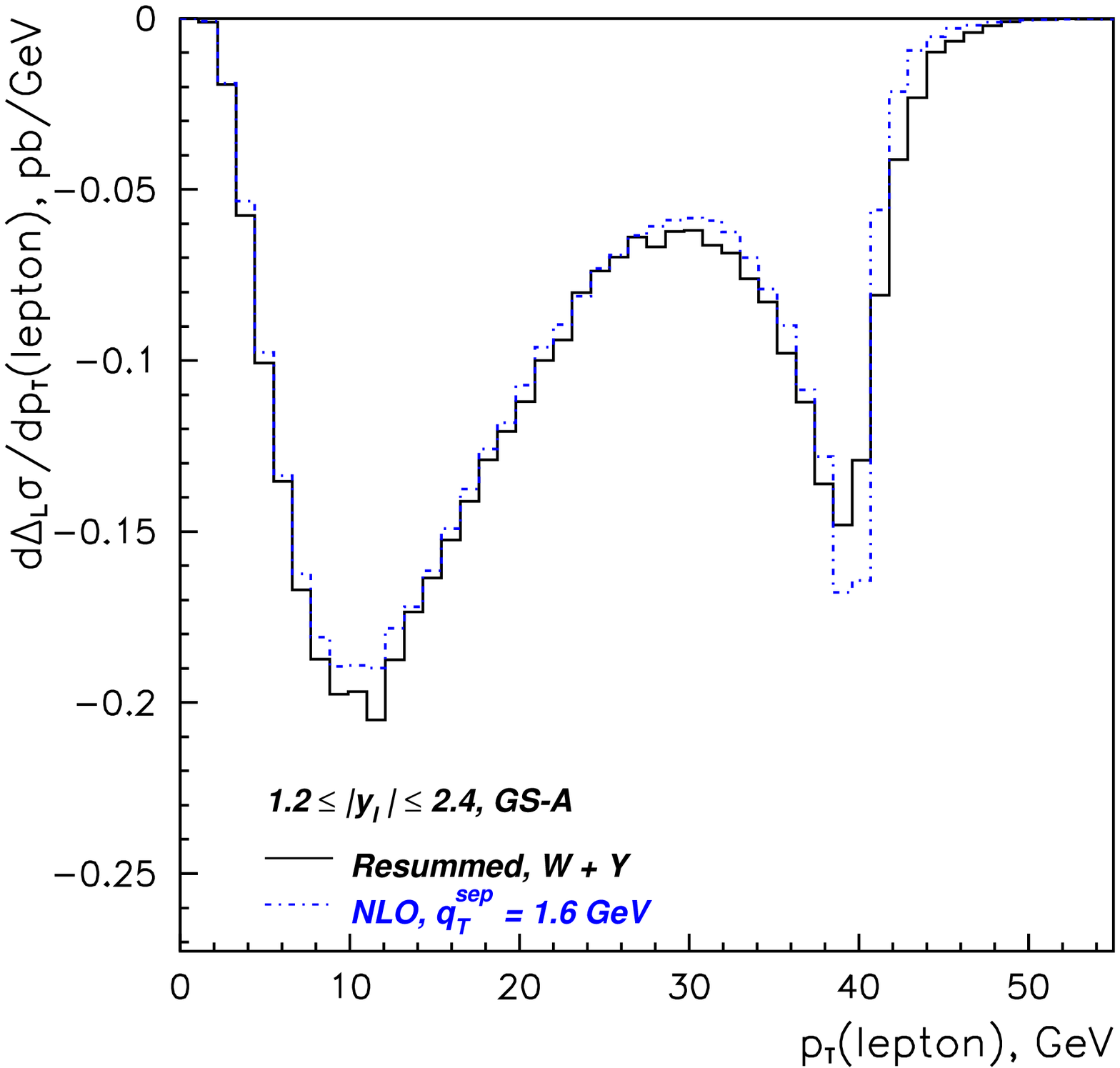}} \subfigure[$-1 \leq y_\ell \leq 1$]{\includegraphics[  width=0.50\textwidth,
  keepaspectratio]{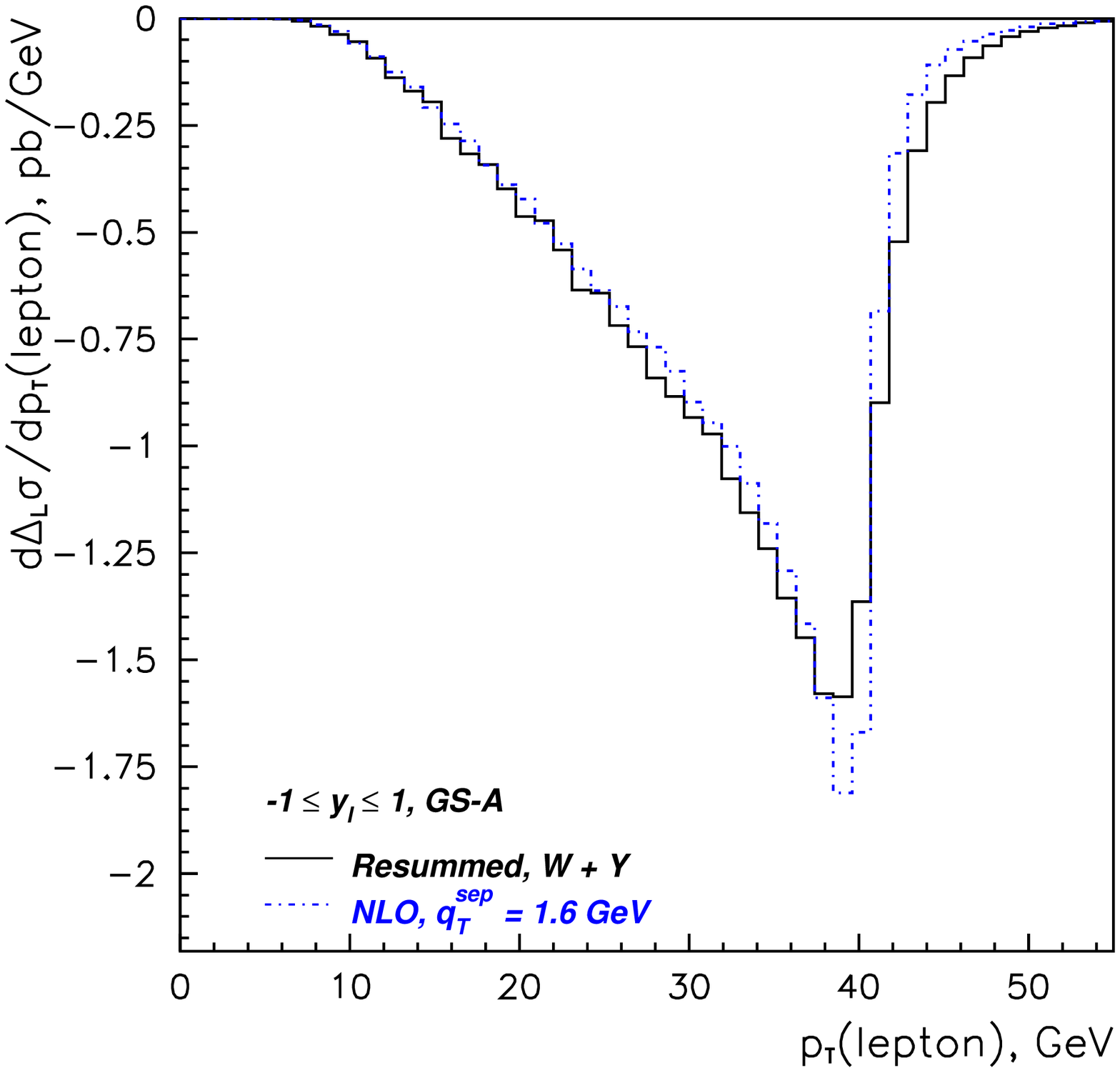}} \end{center}

\caption{\label{fig:ptl_pol_wp}Single-spin resummed (solid) and NLO (dashed)
cross sections $d\Delta _{L}\sigma /dp_{T\ell }$ in the process $\Delta _{L}pp\rightarrow (W^{+}\rightarrow \ell ^{+}\nu _{\ell })X$
for various cuts on the lepton's rapidity. Gehrmann-Stirling PDF set
A \cite{Gehrmann:1996ag} were used. }
\end{figure}

\begin{figure}[p]
\begin{center}\subfigure[No cuts]{\includegraphics[  width=0.50\textwidth,
  keepaspectratio]{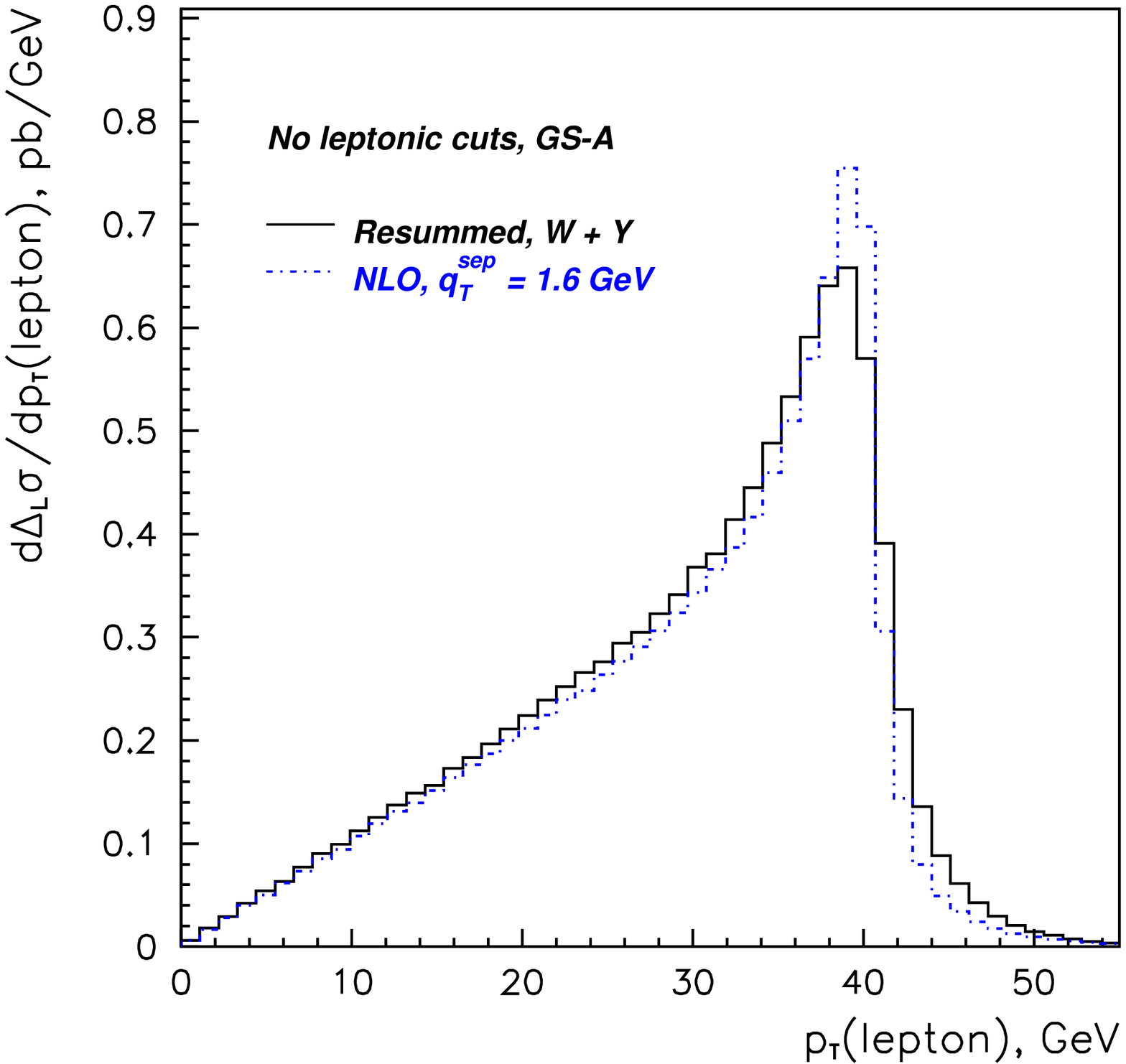}}\subfigure[$1.2\leq |y_\ell |\leq 2.4$]{\includegraphics[  width=0.50\textwidth,
  keepaspectratio]{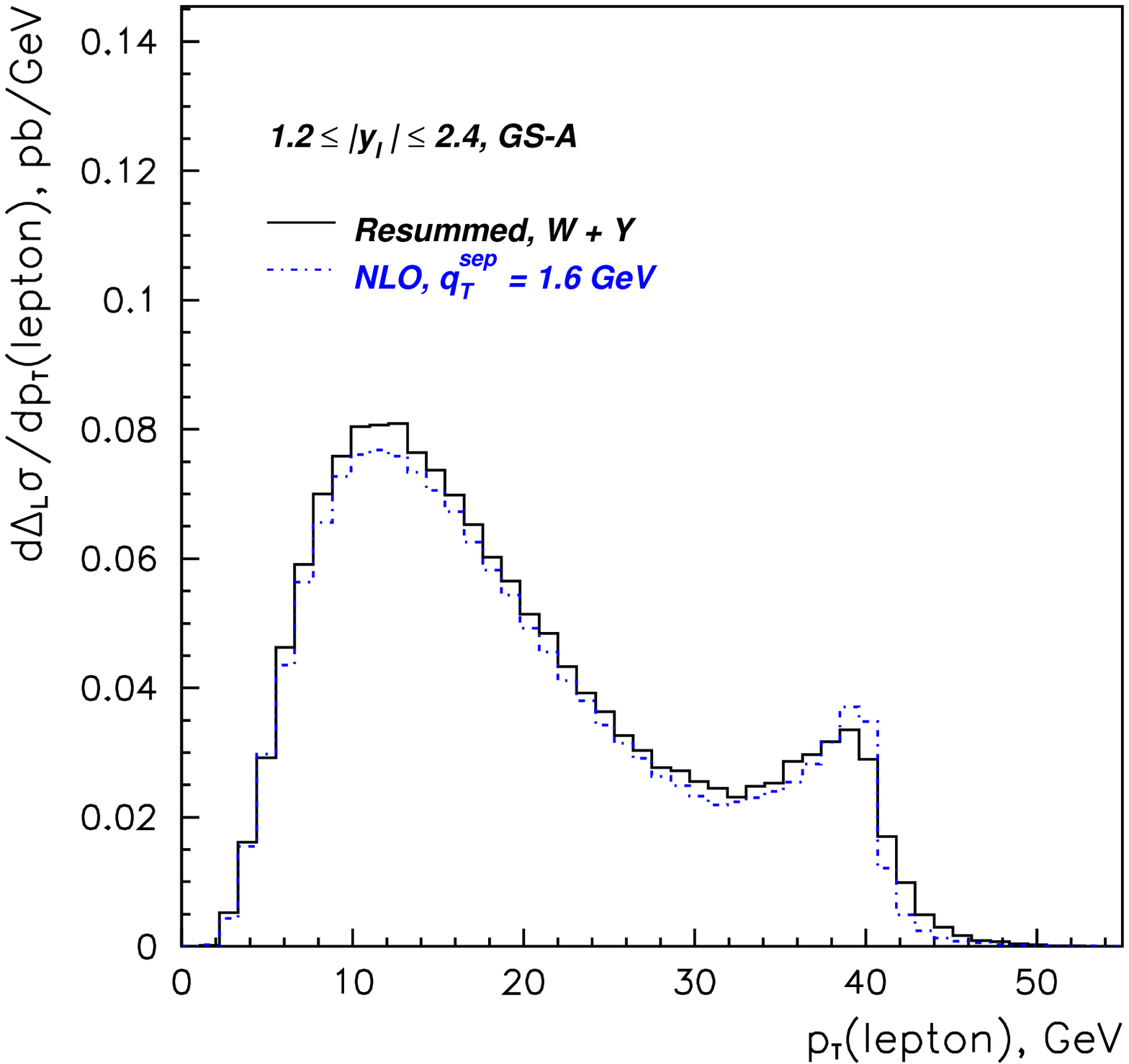}} \subfigure[$-1 \leq y_\ell \leq 1$]{\includegraphics[  width=0.50\textwidth,
  keepaspectratio]{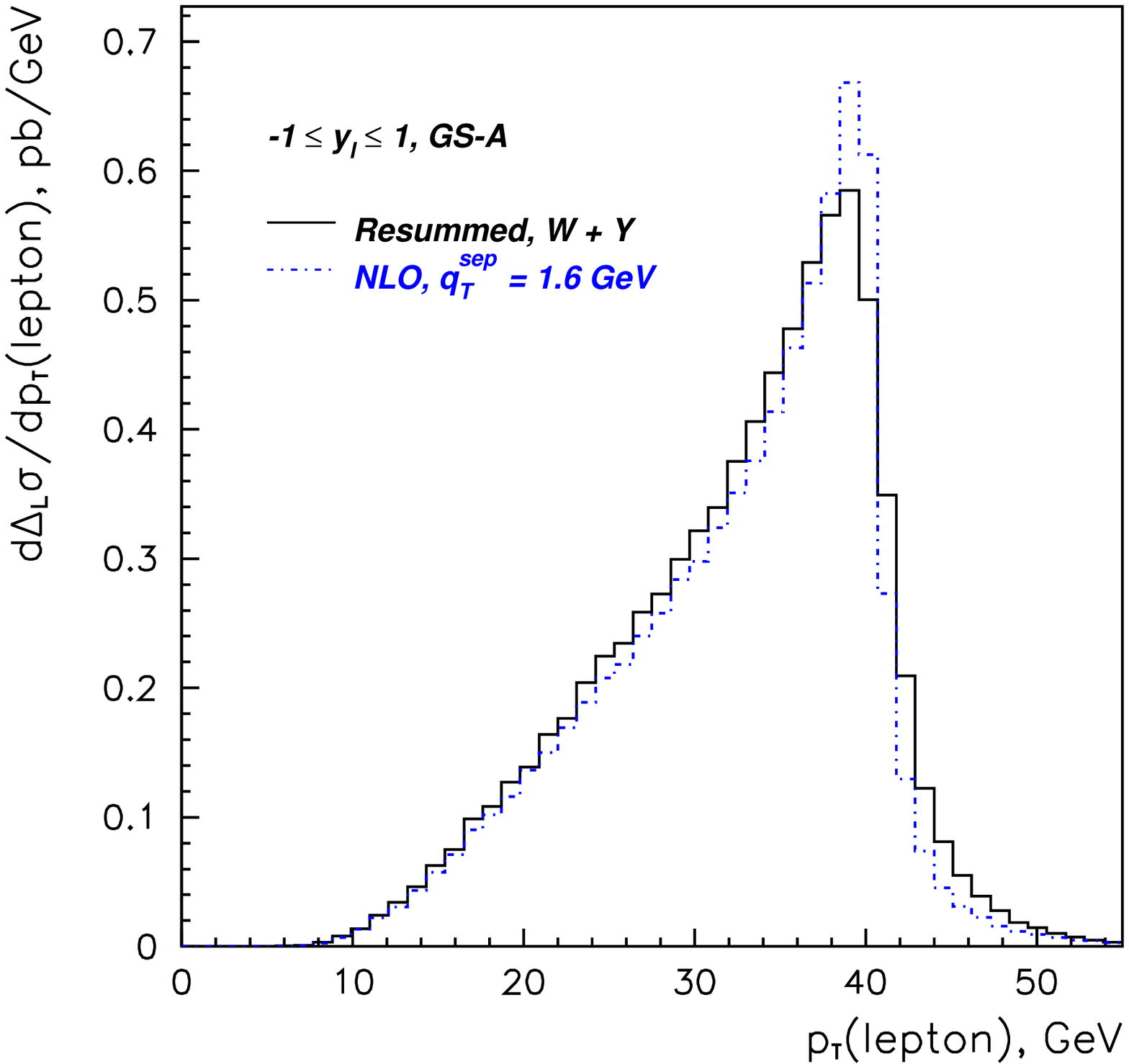}} \end{center}

\caption{\label{fig:ptl_pol_wm}Single-spin resummed (solid) and NLO (dashed)
cross sections $d\Delta _{L}\sigma /dp_{T\ell }$ in the process $\Delta _{L}pp\rightarrow (W^{-}\rightarrow \ell ^{-}\bar{\nu }_{\ell })X$
for various cuts on the lepton's rapidity. Gehrmann-Stirling PDF set
A \cite{Gehrmann:1996ag} was used. }
\end{figure}

The $p_{T\ell }$ distributions in the single-spin $W^{+}$ boson
production are shown in Fig.~\ref{fig:ptl_pol_wp}.  
 The cross sections $d\Delta _{L}\sigma /dp_{T\ell }$ in $W^{+}$
boson production are negative, because the charged weak coupling is
purely left-handed, and the dominant polarized PDF $\Delta u(x)$
is positive in the RHIC range of $x$. Again, we see that the NLO
calculation cannot reliably predict the distribution of $p_{T\ell }$
around and above $M_{W}/2$. For the \PHENIX kinematics, cf.~Fig.~\ref{fig:ptl_pol_wp}(b),
a sharper peak is developed in the distribution of $p_{T\ell }$ around
$M_{W}/2$ as compared to the unpolarized case, cf.~Fig.~\ref{fig:ptl_unp_wp}(b).
The different shape is caused by the different rapidity dependence
in the unpolarized and single-spin cases, as shown in Figs.~\ref{fig:dsigma/dpTldyl}(a)
and \ref{fig:dsigma/dpTldyl}(c). Finally, the distribution $d\sigma /dp_{T\ell }$
for the charged leptons $\ell ^{-}$ in the $W^{-}$ events is shown
in Fig.~\ref{fig:ptl_pol_wm}. %

For the rapidity distributions $d\sigma /dy_{\ell }$ without imposed
cuts on $p_{T\ell }$, the NLO and resummed cross sections are close
to one another, in accordance with the cancellation of soft contributions
in sufficiently inclusive observables. However, according to the
discussion above, the resummation and NLO predictions for $d\sigma /dy_{\ell }$
can be very different if cuts on $p_{T\ell }$ are imposed.

\subsection{Single-spin asymmetries}

\begin{figure}[p]
\begin{center}\includegraphics[  width=1.0\textwidth,
  keepaspectratio]{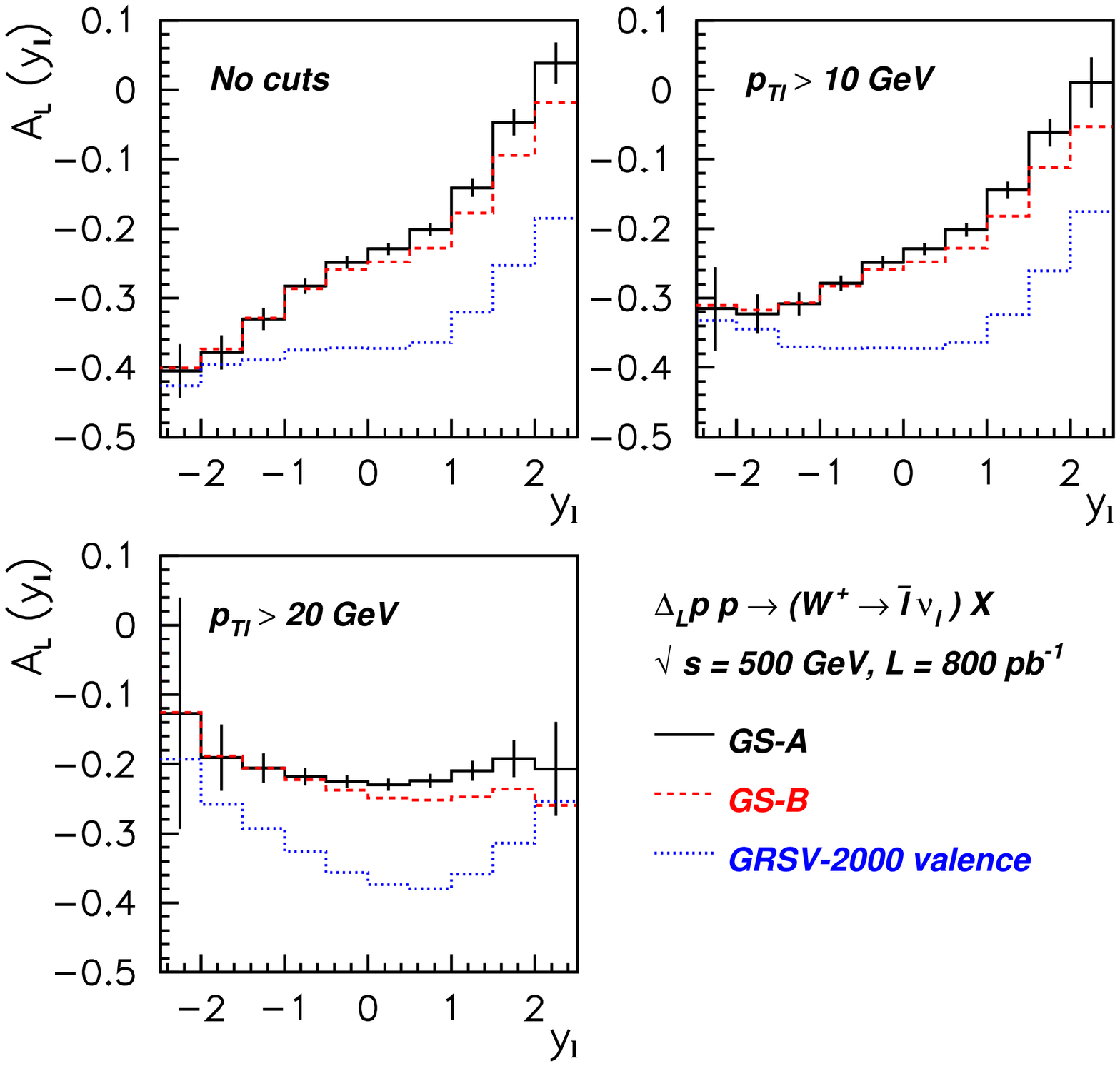}\end{center}

\caption{\label{fig:ALyl_wp} Asymmetries $A_{L}(y_{\ell })$ for various
selection cuts on $p_{T\ell }$ in $W^{+}$ boson production, as predicted
by the resummation calculation. The asymmetry is shown for the Gehrmann-Stirling
PDF sets A (solid) and B (dashed), as well as for the GRSV valence-like
PDF set (dotted). }
\end{figure}

\begin{figure}[p]
\begin{center}\includegraphics[  width=1.0\textwidth,
  keepaspectratio]{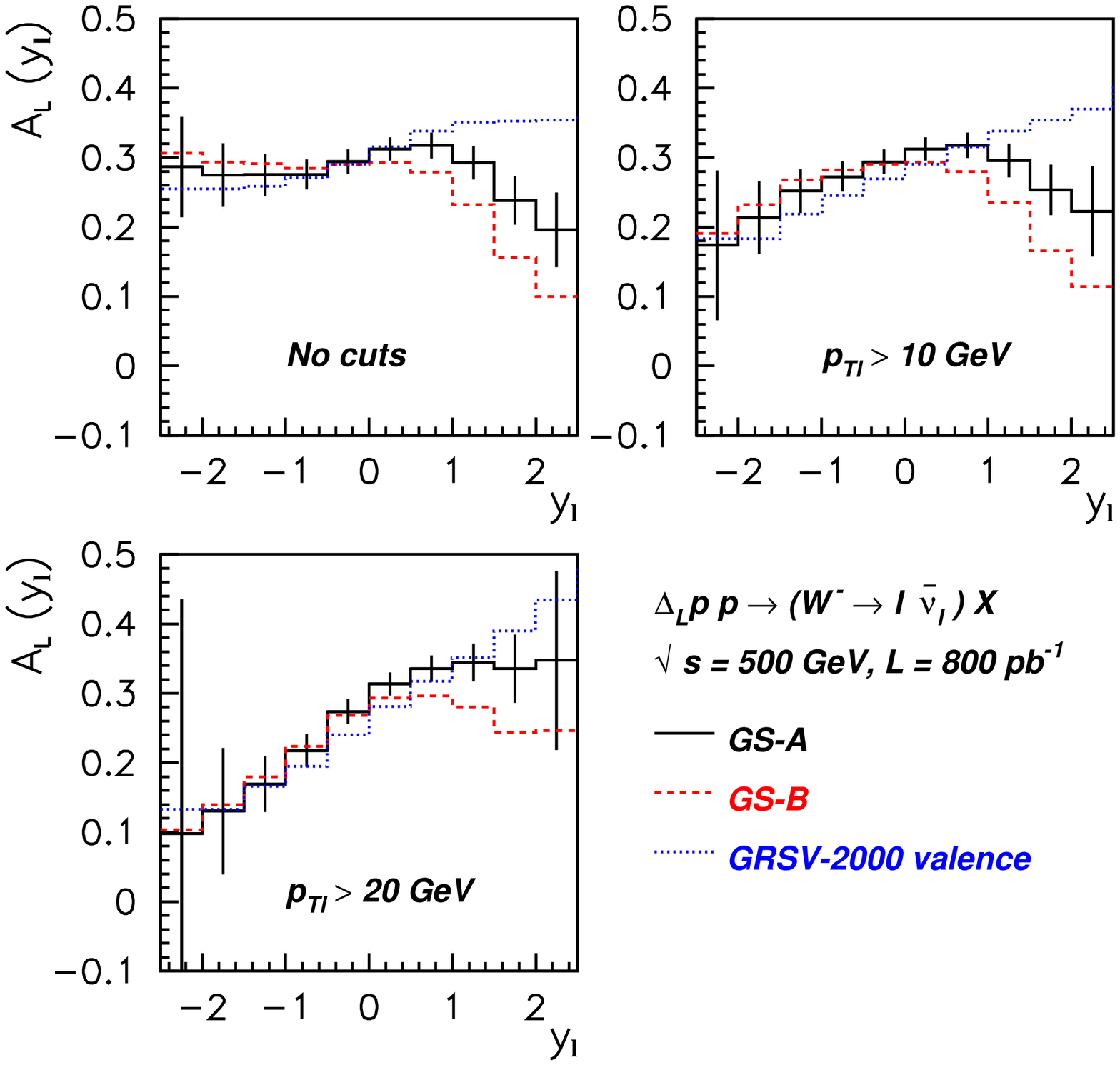}\end{center}

\caption{\label{fig:ALyl_wm} Asymmetries $A_{L}(y_{\ell })$ for various
selection cuts on $p_{T\ell }$ in $W^{-}$ boson production, as predicted
by the resummation calculation. The asymmetry is shown for the Gehrmann-Stirling
PDF sets A (solid) and B (dashed), as well as for the GRSV valence-like
PDF set (dotted). }
\end{figure}

\begin{figure}[p]
\begin{center}\includegraphics[  width=1.0\textwidth,
  keepaspectratio]{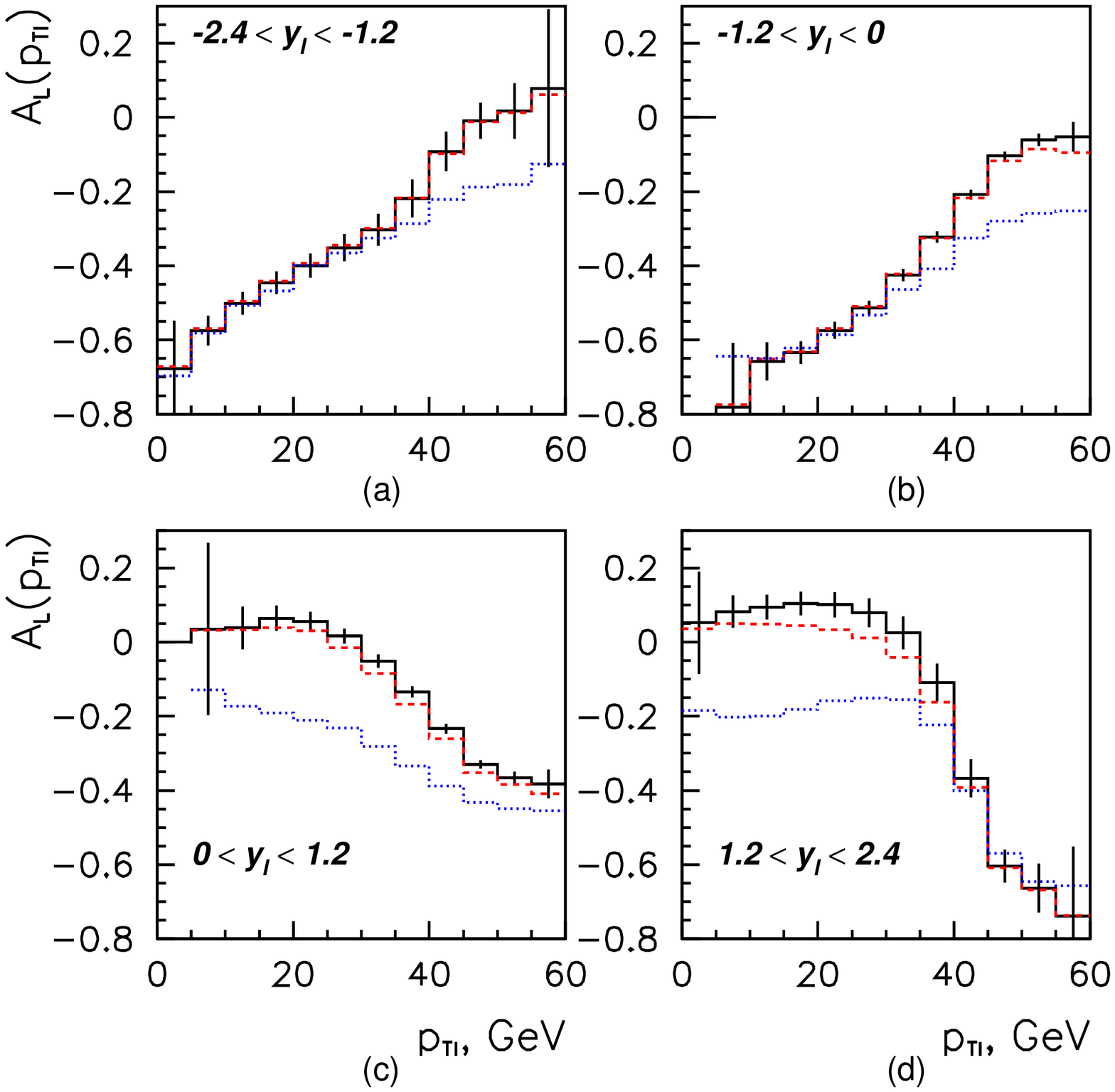}\end{center}

\caption{\label{fig:ALptl_wp} Asymmetries $A_{L}(p_{T\ell })$ for various
selection cuts on $y_{\ell }$ in $W^{+}$ boson production, as predicted
by the resummation calculation. The asymmetry is shown for the Gehrmann-Stirling
PDF sets A (solid) and B (dashed), as well as for the GRSV valence-like
PDF set (dotted). }
\end{figure}

\begin{figure}[p]
\begin{center}\includegraphics[  width=1.0\textwidth,
  keepaspectratio]{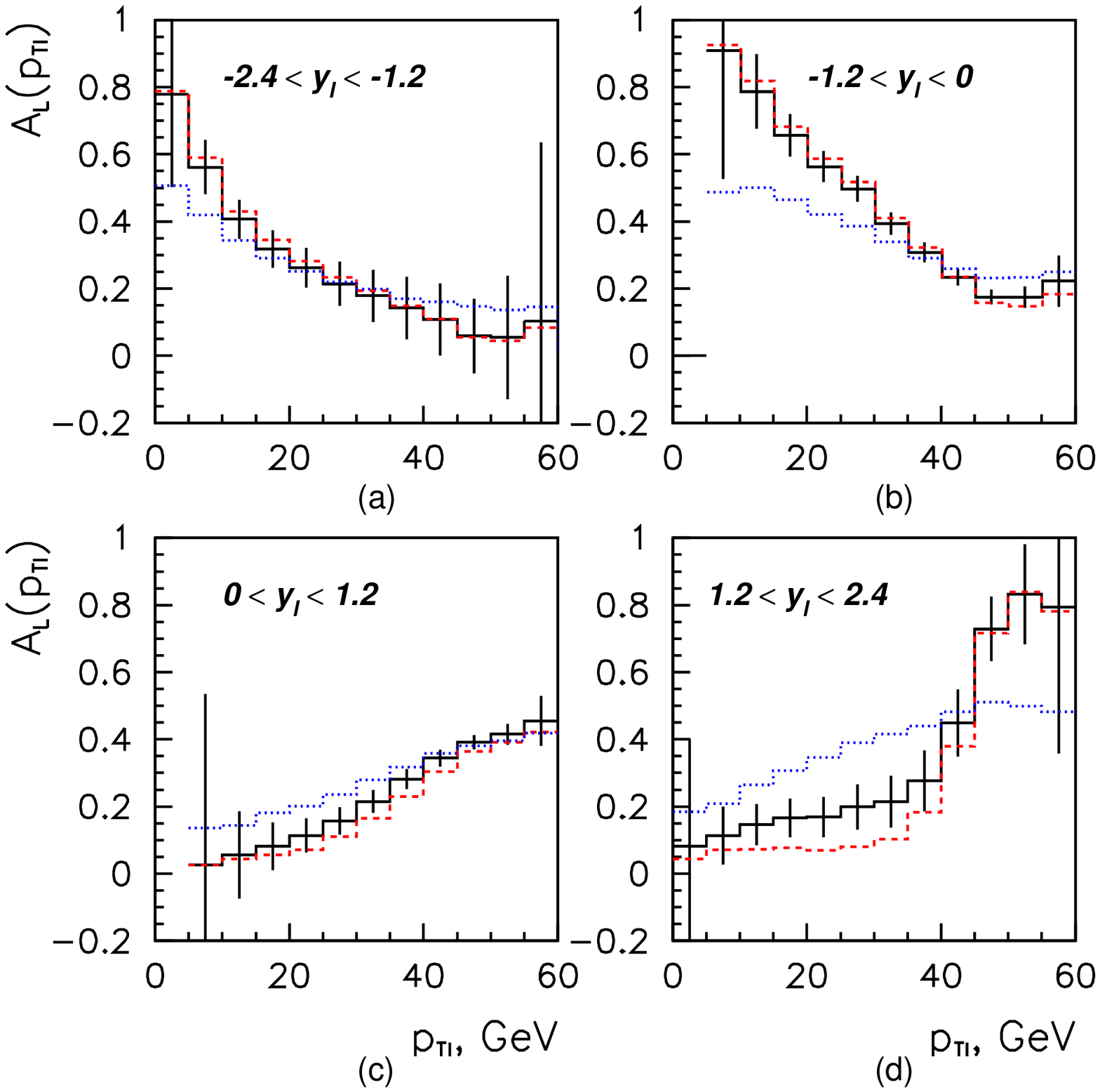}\end{center}

\caption{\label{fig:ALptl_wm} Asymmetries $A_{L}(p_{T\ell })$ for various
selection cuts on $y_{\ell }$ in $W^{-}$ boson production, as predicted
by the resummation calculation. The asymmetry is shown for the Gehrmann-Stirling
PDF sets A (solid) and B (dashed), as well as for the GRSV valence-like
PDF set (dotted). }
\end{figure}

We now turn our attention to the single-spin asymmetry. Since the
only straightforward signature of the $W$ bosons at RHIC is the observation
of the secondary charged leptons, it is important to understand these
asymmetries at the lepton level. In Figs.~\ref{fig:ALyl_wp} and
~\ref{fig:ALyl_wm}, we show the lepton-level asymmetries $A_{L}(y_{\ell })$
in the $W^{+}$ and $W^{-}$ events, respectively, for various cuts
on $p_{T\ell }$. 
In these figures,
the positive rapidity direction is in the moving direction of the
polarized proton beam.
Similarly, Figs.~\ref{fig:ALptl_wp} and \ref{fig:ALptl_wm}
show the asymmetries $A_{L}(p_{T\ell })$ for various cuts on $y_{\ell }$.
All asymmetries are derived using the Gehrmann-Stirling PDF sets A
and B \cite{Gehrmann:1996ag} and GRSV-2000 valence-like PDF set \cite{Gluck:2000dy}
in the resummation calculation. We found by experimenting with other
available PDF sets in Refs.~\cite{Gehrmann:1996ag,Gluck:1996yr,Gluck:2000dy}
that the shown graphs characterize well the possible span in the asymmetry
due to the different polarized PDF parametrizations. The projected
statistical errors are estimated according to Eq.~(\ref{deltaAL})
for the bin sizes shown in the figure. 

According to the figures, the lepton-level asymmetries can be accurately
measured for both $W^{+}$ and $W^{-}$ bosons. These directly observed
asymmetries can efficiently discriminate between different PDF sets;
hence, they provide a viable alternative to the less accessible asymmetry
$A_{L}(y_{W})$. For both $W^{+}$and $W^{-}$ boson samples (cf.~Figs.~\ref{fig:ALyl_wp}
and \ref{fig:ALyl_wm}), the power to discriminate between different
PDF sets is larger in the region of the positive rapidity $y_{\ell }$,
especially for the $W^{-}$ boson data. It is interesting to observe
that, in different ranges of $y_{\ell },$ the largest variations
due to the choice of the PDFs appear in different ranges of $p_{T\ell }$.
In $W^{+}$ boson production, the largest variation in the asymmetries
$A_{L}(p_{T\ell })$ is concentrated at $p_{T\ell }\gtrsim 30-40$
GeV in the backward rapidity region $y_{\ell }<0$ (Fig.~\ref{fig:ALptl_wp}(a)
and (b)) and at $p_{T\ell }\lesssim 30-40$ GeV in the forward rapidity
region $y_{\ell }>0$ (Fig.~\ref{fig:ALptl_wp}(c) and \ref{fig:ALptl_wp}(d)).
In $W^{-}$ boson production the largest variation is observed for
very forward leptons, $1.2<y_{\ell }<2.4$ (Fig.~\ref{fig:ALptl_wm}(d)).
If an additional condition $p_{T\ell }>20$ GeV is imposed, other
regions of $y_{\ell }$ (cf.~Fig.~\ref{fig:ALptl_wm}(a)-\ref{fig:ALptl_wm}(c))
show a smaller sensitivity to different PDF sets. 


\begin{figure}[p]
\begin{center}\includegraphics[  width=1.0\textwidth,
  keepaspectratio]{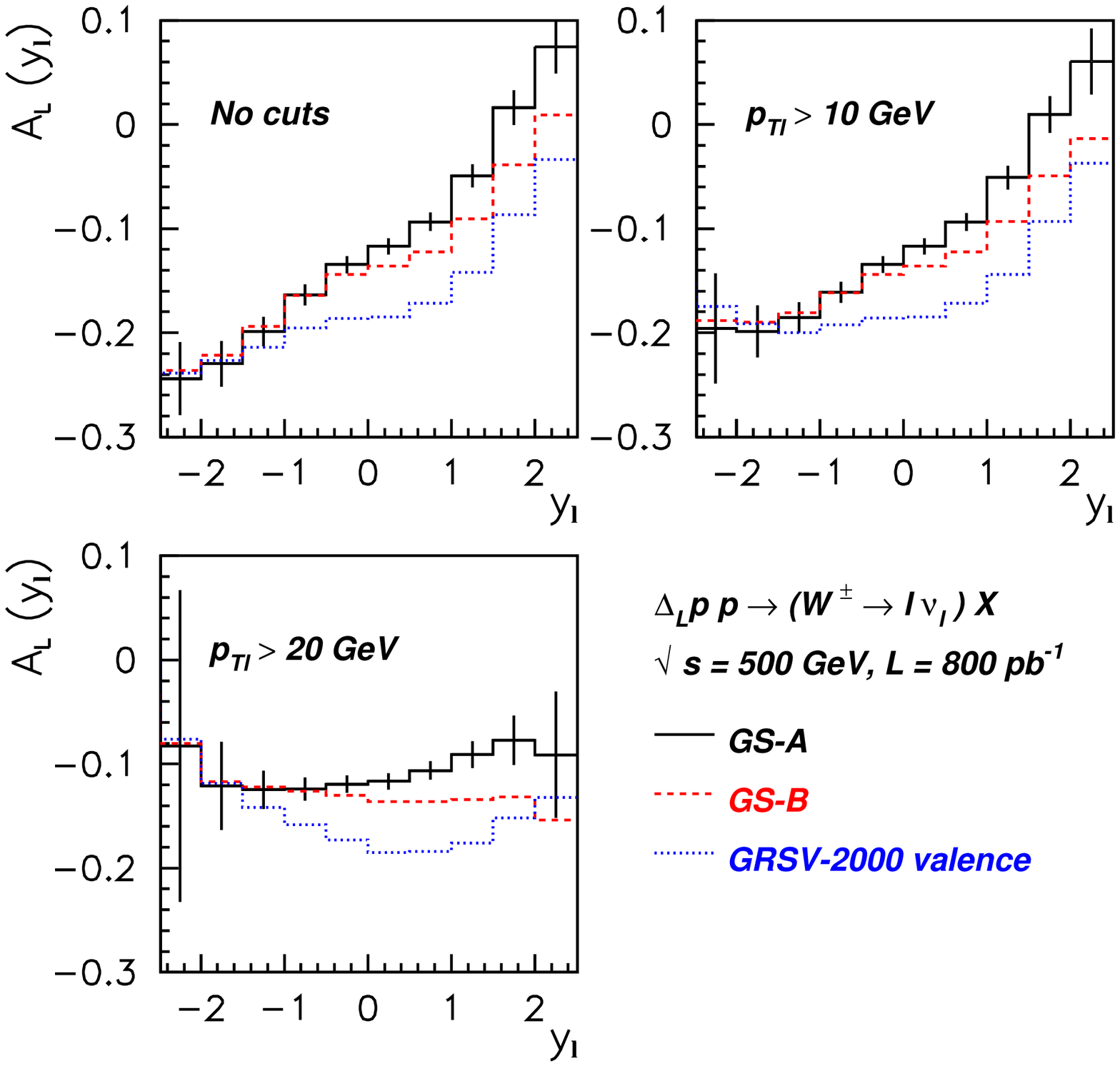}\end{center}

\caption{\label{fig:ALyl_wpwm} Asymmetries $A_{L}(y_{\ell })$ for various
selection cuts on $p_{T\ell }$ for the combined sample of $W^{+}$
and $W^{-}$ bosons, as predicted by the resummation calculation.
The asymmetry is shown for the Gehrmann-Stirling PDF sets A (solid)
and B (dashed), as well as for the GRSV valence-like PDF set (dotted). }
\end{figure}

\begin{figure}[p]
\begin{center}\includegraphics[  width=1.0\textwidth,
  keepaspectratio]{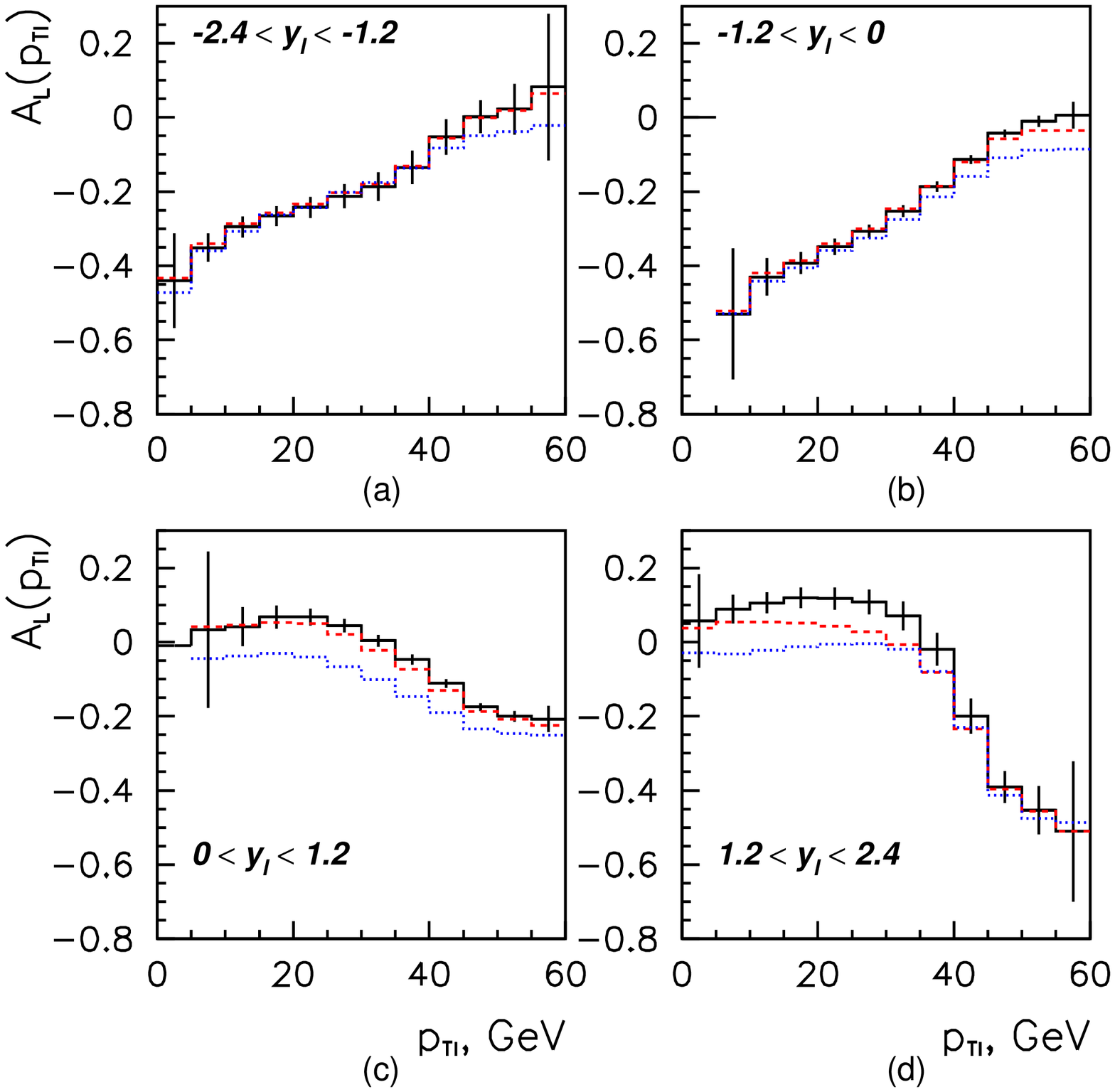}\end{center}

\caption{\label{fig:ALptl_wpwm} Asymmetries $A_{L}(p_{T\ell })$ for various
selection cuts on $y_{\ell }$ for the combined sample of $W^{+}$
and $W^{-}$ bosons, as predicted by the resummation calculation.
The asymmetry is shown for the Gehrmann-Stirling PDF sets A (solid)
and B (dashed), as well as for the GRSV valence-like PDF set (dotted). }
\end{figure}

During the first years of RHIC spin program, it may not be possible
to distinguish between the positively and negatively charged leptons
from the $W$ boson decay, particularly before the solenoid field
of the \STAR detector is fully operational. It is therefore interesting
to examine the possibility of measuring the single-spin asymmetries
in the combined sample of $W^{^{+}}$ and $W^{-}$ bosons, without
distinguishing between charged leptons and anti-leptons. Such asymmetries
are shown in Figs.~\ref{fig:ALyl_wpwm} and \ref{fig:ALptl_wpwm}.
Since there are about three $W^{+}$ bosons produced per each $W^{-}$
boson, these asymmetries are close to the $W^{+}$ boson asymmetries
shown in Figs.~\ref{fig:ALyl_wp} and \ref{fig:ALyl_wm}. Therefore,
these asymmetries will be able to distinguish between different PDF
sets for a sufficiently high number of events.%
\footnote{Of course, the projected error bars in Figs.~\ref{fig:ALyl_wpwm}
and \ref{fig:ALptl_wpwm} (evaluated for ${\mathcal{L}}=800\mbox {\, pb}^{-1}$)
will have to be adjusted to agree with the actual integrated luminosity
at the time of the measurement.%
} However, they cannot completely replace the single-spin asymmetries
for independent $W^{+}$ and $W^{-}$ boson samples, which have better
discriminating power and allow us to distinguish between contributions
of different parton flavors. 


\subsection{Nonperturbative contribution}

As was discussed above, the transverse momentum distributions for
vector bosons are affected by the soft and collinear radiation, which
can be described only by means of all-order resummation. In addition,
the distributions at very small $q_{T}$ are sensitive to nonperturbative
contributions characterized by large impact parameters $b\gtrsim 1\mbox {\, GeV}^{-1}$.
As a result, the shape of the lepton-level distribution $d\sigma /dp_{T\ell }$
around its peak at about $p_{T\ell }=M_{W}/2$ is affected by both
perturbative and nonperturbative QCD radiation, in addition to the
nonzero width of the $W$ boson. While the perturbative soft and collinear
contributions can be calculated order-by-order in the resummation
formalism, the nonperturbative contributions (associated with the
nonperturbative transverse motion of partons inside the proton) can
only be approximated by phenomenological parametrizations. In particular,
based on the factorization properties of the resummed cross sections,
one can hypothesize a simple relationship between the nonperturbative
Sudakov factors in the unpolarized, single-spin, and double-spin vector
boson production:\begin{equation}
\left.S_{NP}\right|_{single-spin}=\frac{1}{2}\left(\left.S_{NP}\right|_{unpolarized}+\left.S_{NP}\right|_{double-spin}\right).\label{SNPcrossing}\end{equation}
Furthermore, $S_{NP}$ is usually assumed to be independent of the
type of the electroweak hard probe, so that it is the same in $\gamma ^{*},$
$W,$ and $Z$ boson production. If the relationship (\ref{SNPcrossing})
for the spin dependence of $S_{NP}$ indeed holds in nature, the nonperturbative
contributions in single-spin $W$ boson production can be determined
unambiguously once the parametrizations of $S_{NP}$ on the right-hand
side of Eq.~(\ref{SNPcrossing}) are available, e.g., from the global
analysis of the distributions $d\sigma /dq_{T}$ in the unpolarized
and double-spin $\gamma ^{*}$ or $Z^{0}$ production. 

\begin{figure}[p]
\begin{center}\includegraphics[  width=0.80\textwidth,
  keepaspectratio]{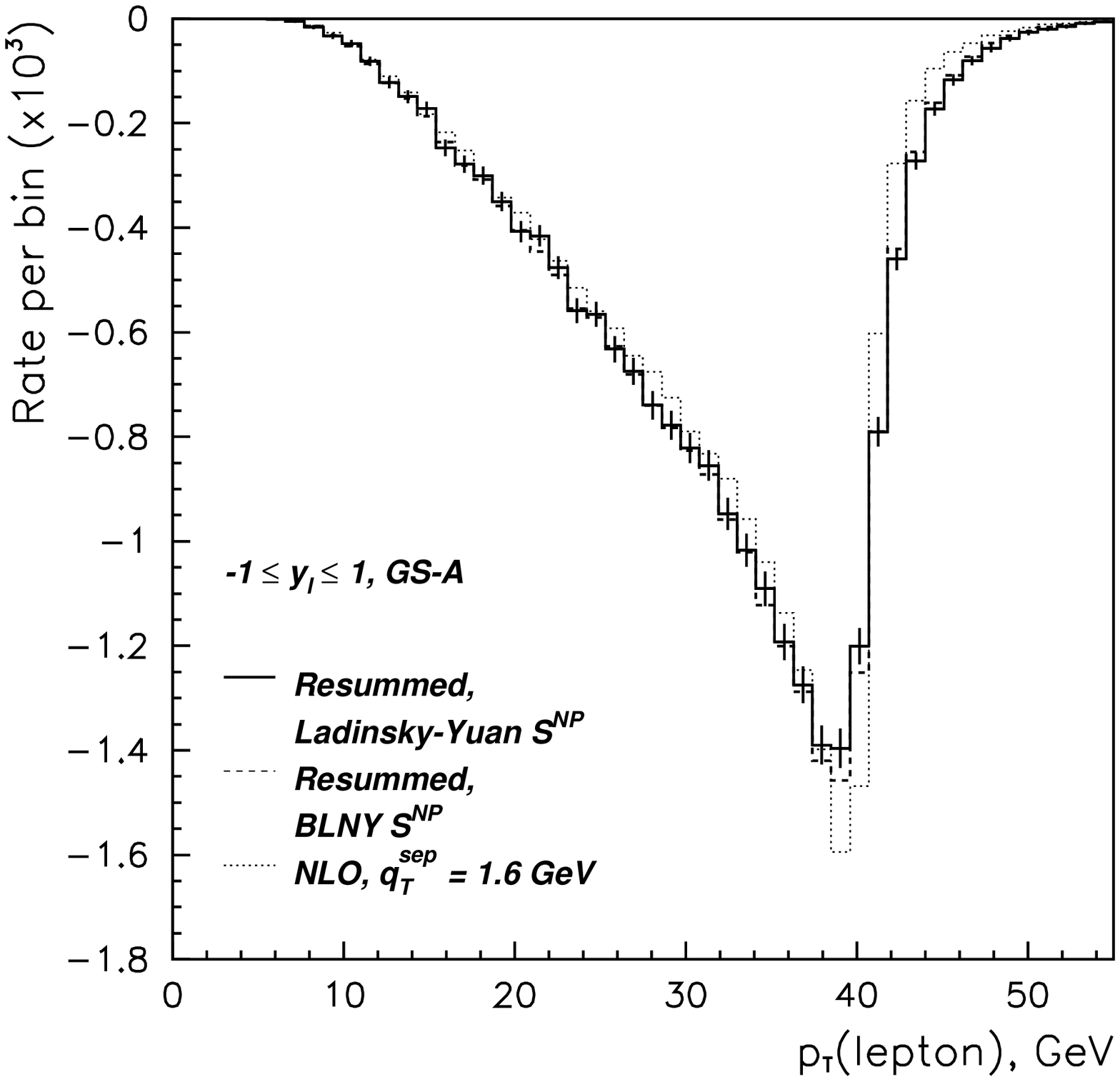}\end{center}

\caption{\label{fig:ptl}The single-spin charged lepton transverse momentum
distribution $d\Delta _{L}\sigma /dp_{T\ell }$ for $W^{+}$ boson
production discussed in the main text. The nonperturbative parts of
the resummed cross sections were approximated by the Ladinsky-Yuan
\cite{Ladinsky:1994zn} (solid) and Brock-Landry-Nadolsky-Yuan \cite{Landry:2002ix}
(dashed) phenomenological parameterizations from unpolarized vector
boson production. The NLO cross section is shown for comparison as
a dotted line. The distributions are calculated using the Gehrmann-Stirling
PDF set A \cite{Gehrmann:1996ag}.}
\end{figure}

It is well known from the unpolarized studies (see, e.g., Refs.~\cite{CSS,DWS,Arnold:1991yk})
that the importance of the nonperturbative contributions in the massive
boson production at high-energy colliders is strongly reduced as compared
to Drell-Yan pair production in fixed-target experiments. Nonetheless,
this dependence is not entirely negligible at small transverse momenta
of $W$ bosons, roughly for $q_{T}\sim 5$ GeV or less. To give an
idea about the typical size of the nonperturbative contributions,
Fig.~\ref{fig:ptl} shows the event rate for the single-spin cross
section $d\Delta _{L}\sigma /dp_{T\ell }$ at ${\mathcal{L}}=800\mbox {\, pb}^{-1}$,
integrated over the \STAR acceptance range $-1\leq y_{\ell }\leq 1$.
This rate was calculated using two parameterizations \cite{Ladinsky:1994zn,Landry:2002ix}
of the nonperturbative part, which were found from the analysis of
unpolarized vector boson production. Correspondingly, the resummed
cross sections were calculated with the assumption that $\left.S_{NP}\right|_{single-spin}=\left.S_{NP}\right|_{unpolarized}$.
It can be seen that the sensitivity to the nonperturbative input is
small, but, nonetheless, visible near the Jacobian peak. For comparison,
we also included the NLO cross section calculated using the phase
space slicing method. The NLO curve substantially deviates from the
resummation curves, and, moreover, its shape can be drastically modified
by varying the phase space slicing parameter $q_{T}^{sep}$. In contrast,
practically all features of the resummation curve are determined unambiguously
by the perturbative contribution, and the remaining small dependence
on the nonperturbative contributions is pinned down by utilizing a
phenomenological parameterization discussed above. Needless to say,
RHIC should explore the spin properties of nonperturbative contributions
by studying distributions $d\sigma /dq_{T}$ in the double-spin $\gamma ^{*}$
and $Z^{0}$ production and $d\sigma /dp_{T\ell }$ in single-spin
and double-spin $W$ boson production. %

\section{Conclusion}

Due to the maximal parity violation in the weak current coupling,
the longitudinal single-spin asymmetry of $W$ boson production is
strongly sensitive to the polarized parton distributions. At the center-of-mass
energy of 500 GeV, $W$ boson production events at RHIC can clearly
discriminate between various polarized quark luminosities. As a $pp$
collider, RHIC probes the sea quark PDFs in the region of relatively
high $x,$ where such PDFs are not adequately constrained even in
the unpolarized case. Therefore, RHIC experiments also have a good
potential to reduce the uncertainties in the unpolarized sea quark
PDFs in a dedicated study of spin-averaged cross sections.

Since \PHENIX and \STAR detectors do not determine the missing transverse
energy associated with the neutrino from $W$ boson decay, and because
the charged leptons from the decay can only be observed in a part
of the $4\pi $ solid angle, the reconstruction of the asymmetries
at the level of $W$ bosons is not easy to realize. As an alternative
to the asymmetry $A_{L}(y_{W})$ for the $W$ boson rapidity distribution
$d\sigma /dy_{W}$, we advocate the measurement of the asymmetries
$A_{L}(y_{\ell }$) and $A_{L}(p_{T\ell })$ for the distributions
$d\sigma /dy_{\ell }$ and $d\sigma /dp_{T\ell }$ in the rapidity
$y_{\ell }$ and transverse momentum $p_{T\ell }$ of the secondary
charged lepton. These lepton-level asymmetries can be directly observed
at RHIC detectors, they are sizeable, and they can be used to determine
the polarized PDFs with high precision.

To reliably predict the distributions of the charged leptons in all
available phase space, a theoretical calculation is needed for all-order
summation of large logarithmic contributions corresponding to $W$
bosons with a small transverse momentum. At the lepton level, the
largest effect of the logarithm resummation is observed in the distributions
$d\sigma /dp_{T\ell }$ and $d\Delta _{L}\sigma /dp_{T\ell }$ in
the proximity of the Jacobian peak ($p_{T\ell }\approx M_{W}/2$).
The importance of the resummation can be verified by observing a large
difference between the resummed and NLO cross sections near the Jacobian
peak, which happens only when the effect of soft gluon radiation is
important. 

In this paper, we apply the state-of-art resummation calculation \cite{PolWTheory}
for spin-dependent vector boson production to predict reliably the
distributions of the observed leptons from $W$ boson decay. In addition
to the spin-averaged and single-spin cross sections, we presented
a detailed study of single-spin asymmetries for the decay lepton in
the presence of kinematical cuts imposed by \PHENIX and \STAR detectors.
Furthermore, we propose to study the distribution $d\Delta _{L}\sigma /dp_{T\ell }$
at $p_{T\ell }\approx M_{W}/2$ to learn about the spin (in)dependence
of nonperturbative contributions in the CSS resummation formalism.
The study of such contributions for different beam polarizations and
bosons of different types ($\gamma ^{*},$$W^{\pm }$, $Z^{0}$) will
shed light on basic properties of QCD factorization and provide important
clues about the nature of the intrinsic transverse motion of partons
inside the proton.

\section*{Acknowledgments}

Authors would like to thank C.~Balazs, D.~Boer, G.~Bunce, M.~Grosse
Perdekamp, S.~Gupta, J. Kiryluk, N.~Saito, M.~Stratmann, W.~Vogelsang,
and members of the CTEQ collaboration for helpful discussions. We
are grateful to F.~Olness for his comments on the manuscript. We
thank the organizers of RHIC Spin workshops, where preliminary results
of this work were presented. The work of P. M. N. has been supported
by the U.S. Department of Energy and Lightner-Sams Foundation. The
research of C. P. Y. has been supported by the National Science Foundation
under grant PHY-0100677.

\end{document}